\documentclass[useAMS,usenatbib]{mn2e}

\usepackage{amsmath}
\usepackage{graphicx}
\usepackage{color}
\usepackage{booktabs}

\usepackage{hyperref}
\hypersetup{colorlinks,citecolor=blue,linkcolor=blue,urlcolor=blue}

\def\lsim{~\rlap{$<$}{\lower 1.0ex\hbox{$\sim$}}}

\def\gsim{~\rlap{$>$}{\lower 1.0ex\hbox{$\sim$}}}

\def\HIItext{H\,\textsc{ii}}

\def\Meraxes{\textsc{Meraxes}}

\def\tocf{\textsc{21cmfast}}

\newcommand{\vect}[1]{\boldsymbol{#1}}

\newcommand{\ra}[1]{\renewcommand{\arraystretch}{#1}}

\title[DRAGONS: Morphology and statistics of reionization]{Dark-ages~reionization~\&~galaxy~formation~simulation~V: morphology and statistical signatures of reionization}

\author[P.~M.~Geil et al.]
{\parbox{\textwidth}{
Paul~M.~Geil$^{1}$\thanks{E-mail: geil.p@unimelb.edu.au},
Simon~J.~Mutch$^{1}$,
Gregory~B.~Poole$^{1}$,
Paul~W.~Angel$^{1}$,
Alan~R.~Duffy$^{2}$,
Andrei Mesinger$^{3}$ and
J.~Stuart~B.~Wyithe$^{1}$}\vspace{0.4cm}\\
\parbox{\textwidth}{$^{1}$School of Physics, The University of Melbourne, Parkville, Victoria 3010, Australia\\
$^{2}$Centre for Astrophysics and Supercomputing, Swinburne University of Technology, PO Box 218, Hawthorn, VIC 3122, Australia\\
$^{3}$Scuola Normale Superiore, Piazza dei Cavalieri 7, I-56126 Pisa, Italy}}

\begin{document}

\date{\today}

\pagerange{\pageref{firstpage}--\pageref{lastpage}} \pubyear{2014}

\maketitle

\label{firstpage}

\begin{abstract}
We use the Dark-ages, Reionization And Galaxy-formation Observables from Numerical Simulations (DRAGONS) framework to investigate the effect of galaxy-formation physics on the morphology and statistics of ionized hydrogen (\HIItext) regions during the Epoch of Reioinization (EoR). DRAGONS self-consistently couples a semi-analytic galaxy-formation model with the inhomogeneous ionizing UV background, and can therefore be used to study the dependence of morphology and statistics of reionization on feedback phenomena of the ionizing source galaxy population. Changes in galaxy-formation physics modify the sizes of \HIItext~regions and the amplitude and shape of 21-cm power spectra. Of the galaxy physics investigated, we find that supernova feedback plays the most important role in reionization, with \HIItext~regions up to $\approx 20$~per~cent smaller and a fractional difference in the amplitude of power spectra of up to $\approx 17$~per~cent at fixed ionized fraction in the absence of this feedback. We compare our galaxy-formation-based reionization models with past calculations that assume constant stellar-to-halo mass ratios and find that with the correct choice of minimum halo mass, such models can mimic the predicted reionization morphology. Reionization morphology at fixed neutral fraction is therefore not uniquely determined by the details of galaxy formation, but is sensitive to the mass of the haloes hosting the bulk of the ionizing sources. Simple EoR parametrizations are therefore accurate predictors of reionization statistics. However, a complete understanding of reionization using future 21-cm observations will require interpretation with realistic galaxy-formation models, in combination with other observations.

\end{abstract}


\section{Introduction}
\label{Introduction}

Many fundamental questions about the first galaxies remain unanswered despite recent progress in both observations and theoretical studies. Future indirect observations of the first sources of light, through their impact on the surrounding intergalactic medium (IGM), promise to shed new light on some of the important processes involved in their formation and evolution. For this reason, the period during which galaxies reionized the Universe -- the Epoch of Reionization (EoR) -- has become a focus of both theory and observation \citep*[see][for reviews]{FCK2006,FOB2006,MW2010}.

Observations of the cosmic microwave background (CMB) and high-redshift sources (such as quasars, Lyman-break galaxies, Lyman-$\alpha$ emitters and gamma-ray bursts) have allowed some constraints to be placed on the timing and duration of the EoR. Without further measurements of the degree of reionization throughout this period, however, its detailed history will remain unknown. Furthermore, the constraints provided by these observations are confined to the global ionization state of the IGM, with limited information on the sources and environmentally-dependent astrophysics at play. The new observational window provided by low-frequency radio experiments will be the first direct probe of neutral hydrogen during reionization. Together, instruments such as LOFAR\footnote{\url{http://www.lofar.org}}, MWA\footnote{\url{http://www.mwatelescope.org}}, PAPER\footnote{\url{http://eor.berkeley.edu}} and GMRT\footnote{\url{http://gmrt.ncra.tifr.res.in}} promise to yield statistical measurements of the state and morphology of cosmic hydrogen throughout the EoR through fluctuations in the redshifted 21-cm emission line intensity. The next generation of instruments, including SKA\footnote{\url{http://www.skatelescope.org}} and HERA\footnote{\url{http://reionization.org}}, will allow high-resolution tomographic images of the ionized structure around individual regions of ionized hydrogen (\HIItext) to be made \citep[see, e.g.,][]{MELLEMA2015,WGK2015}.

Combined with detailed simulations, these observations should pave the way to a better understanding of the connection between galaxies and reionization. Numerous theoretical methods have been used to study reionization. Analytic modelling \citep[e.g.][]{SHAPIRO1994,WL2003,FZH2004} can provide insight into some of the processes involved, but cannot include many of the most important physical effects, and do not generate spatial realizations. N-body simulations, post-processed using radiative transfer methods based on ionizing sources placed within dark matter haloes, provide a description of the structure, but not a self-consistent calculation of source properties \citep[see, e.g.,][]{ILIEV2005,MLZD2007}. Hydrodynamical simulations that include radiative transfer \citep[see, e.g.,][]{GNEDIN2000a,PETKOVA2011,PAARD2013,BAUERetal2015} include the effects of feedback and follow structure formation and cosmic reionization self-consistently but do so at great computational expense when performed with the temporal cadence, cosmic volume and resolution required to capture the evolution and relevant spatial scales of reionization\footnote{Simulations on scales of $\sim 100$~Mpc have been found to be large enough to adequately capture both the structure and duration of reionization. Additional power, however, is found to come from fluctuations on larger scales due to the clustering of ionized regions and therefore simulations on scales of at least $\sim 500$~Mpc are required to fully capture this.}. As a compromise, semi-numerical simulations \citep{ZAHN2005,MF2007,GW2008,ALV2009,CHR2009,THOMAS2009,MFC2011} are computationally inexpensive and provide an efficient means of exploring high-dimenisional parameter spaces and very large volumes (and hence rare objects). However, these simulations have not previously included realistic galaxy formation in their formulation.

Previous theoretical work constraining properties of the ionizing sources throughout the EoR with 21-cm observations has been carried out using a number of approaches. \cite{BARKANA2009} fitted properties of the galaxy population during reionization (such as mean halo mass of the ionizing sources) to simulated 21-cm power spectra using analytical modelling. Extensive investigation using radiative transfer simulations run on top of suites of N-body simulations has been performed by \cite{MLZD2007} and \cite{ILIEV2012}. This work investigated whether or not observations will be able to distinguish reionization due to sources within populations of atomically-cooled haloes of different mass ranges, as well as the effects of ionizing source efficiencies and self-regulation. Other work has explored the effect and relative dominance of supernova feedback during reionization \citep[see, e.g.,][]{KIM2013a}, and the ability of observations to constrain galaxy formation through the effect of variation in the mass- and redshift-dependent escape fraction of ionizing photons \citep{KIM2013b}. Using a Markov chain Monte Carlo analysis tool, \cite{GREIG2015} made estimates of astrophysical parameter contraints from simulated observations, including the ionizing efficiency and mean free path of ionizing photons, as well as lower limits on the virial temperature of star-forming haloes.

The Dark-ages, Reionization And Galaxy-formation Observables from Numerical Simulations (DRAGONS\footnote{\url{http://dragons.ph.unimelb.edu.au}}) project takes a hybrid approach by integrating a semi-numerical calculation of reionization within a semi-analytic model (SAM) of galaxy formation built upon an N-body simulation specifically designed for EoR studies. DRAGONS aims to answer the following questions about galaxy formation and cosmic reionization. What was the nature of the ionizing sources and the processes involved in reionization? What was the relative impact of these processes on cosmic reionization and subsequent galaxy formation? Was reionization a self-regulatory process? Will future observations enable us to distinguish these processes, quantify them, or rule out certain reionization scenarios?

This paper focusses on the morphology and statistical signature of cosmological reionization and is the fifth in a series describing the DRAGONS project. The first paper in the series \citep[][hereafter Paper-I]{DRAGONS1} introduces the collisionless N-body simluation, \emph{Tiamat}, which is used as the basis of this work. The structure of the high-redshift dark matter haloes contained within \emph{Tiamat} is described in the second paper \citep[][Paper-II]{DRAGONS2}. The third paper \citep[][in press, Paper-III]{DRAGONS3} presents \Meraxes, the semi-analytic model of galaxy formation and evolution developed for DRAGONS, while the fourth \citep[][submitted, Paper-IV]{DRAGONS4} discusses the resulting galaxy luminosity functions.

This paper is structured as follows. We begin by describing our simulation and modelling methodologies in Section~\ref{Simulation and Modelling}. In particular, Section~\ref{Simulations} provides a description of the model runs we perform in order to explore the effect of key astrophysical mechanisms included in our simulations. We then present our results in Section~\ref{Results}, where we compare our model runs using various statistics. We discuss aspects of these results in Section~\ref{Discussion} before presenting a summary in Section~\ref{Summary and conclusions}. We include an appendix containing supporting material demonstrating the spatial convergence of our results. All globally-averaged quantities (e.g. neutral fraction) are volume weighted, and distances given in comoving units unless stated otherwise. Our choice of cosmology throughout is the standard spatially-flat \textit{Planck} $\Lambda$CDM cosmology \citep{PLANCK2015} $(h, \Omega_{\rm{m}}, \Omega_{\rm{b}}, \Omega_\Lambda, \sigma_8, n_{\rm{s}})$ = $(0.678, 0.308, 0.0484, 0.692, 0.815, 0.968)$.

\section{Simulation and Modelling}
\label{Simulation and Modelling}

In this section we summarize our simulation and modelling methodologies, beginning with our collisionless N-body simulation (\emph{Tiamat}), then our semi-analytic model (\Meraxes) and finally how we integrate these with the semi-numerical reionization code \tocf~\citep{MFC2011}.

\subsection{Collisionless N-body simulation}
\label{Collisionless N-body simulation}

The collisionless N-body simluation used as the basis of this work is called \emph{Tiamat}. It consists of 2160$^3$ particles in a 100~Mpc, periodic, cubed box evolved using \textsc{gadget}-\footnotesize{2} \citep{GADGET22005}. Initial conditions were generated at $z = 99$ and the simulation was run down to $z = 5$ providing 100 snapshots of particle data equally spaced in time between redshifts 5--35 (roughly one every 11 Myr). Halo finding was performed using \textsc{subfind} \citep{SUBFIND2001} and merger trees from the resulting halo catalogues were created using the methodology to be presented in Poole~et~al.~(in preparation). A triangular-shaped cloud mass assignment scheme \cite[see, e.g.,][]{CUI2008} was used to create density grids used by \Meraxes/\tocf. A complete description of \emph{Tiamat}, resulting halo mass functions and analysis is given in Paper-I.

\subsection{Semi-analytic modelling}
\label{Semi-analytic modelling}

Semi-analytic models parametrize the physics of galaxy formation to enable fast and accurate realizations of galaxy properties within cosmic volumes. The semi-analytic model of galaxy formation used in this work is called \Meraxes, and includes baryonic infall, cooling, star formation, reionization and supernova feedback, metal enrichment, stellar mass recycling, mergers, and ionizing flux from galaxies hosted by dark matter haloes temporarily unresolved in our input merger trees. Calibration was performed so as to agree with observations of the CMB optical depth to electron scattering, and to replicate the observed evolution of the galaxy stellar mass function between redshifts 5--7. A complete description of \Meraxes~is given in Paper-III. A feature of \Meraxes~that is important for its application to reionization is that it is run on `horizontal' merger trees, which enable calculation of the evolution of all galaxies at each time-step, so that feedback processes from neighbouring sources can be incorporated self-consistently.

\subsection{Self-consistent reionization}
\label{Self-consistent reionization}

We simulate cosmic reionization by applying the \tocf~algorithm \citep[details of \tocf~are described in][]{MFC2011} within the \Meraxes~semi-analytic model including the effect of a local/inhomogeneous ionizing UV background (UVB) as described by \cite{SM2013a}. Coupling the galaxy properties modelled by \Meraxes~with the UVB calculated by \tocf~provides a self-consistent, UVB-regulated realization of reionization. This is implemented in the following manner:

\begin{enumerate}
\item At the end of each snapshot, \Meraxes~constructs halo mass, stellar mass and star-formation rate grids;
\item \tocf's excursion-set filtering algorithm calculates the corresponding ionization state and inhomogeneous UVB-intensity grids;
\item \Meraxes~keeps track of the redshift at which each voxel was first ionized, $z_{\rm ion}$, and calculates a baryon fraction modifier, $f_{\rm mod}$, of each halo which is then used to calculate the effect of the local UVB on the amount of infalling bayonic matter available to accrete, cool and form stars for each halo;
\item \Meraxes~evolves all of the galaxies in the simulated volume forward to the next time-step, and then the process is repeated.
\end{enumerate}

Reionization was simulated on a 512$^3$ grid, which we find sufficiently resolves behaviour for analysis in this paper (see Appendix~\ref{App:CT} which discusses the spatial convergence of our results). We now explain key aspects of these steps in more detail, starting with the reionization condition, followed by the incorporation of an inhomogeneous UVB, and finally the feedback and coupling between reionization and galaxy formation.

\subsubsection{Reionization condition}
\label{Reionization condition}

The basic methodology of the \tocf~algorithm is to utilize an excursion-set approach to identify \HIItext~regions and provide a neutral hydrogen fraction ($x_{\rm H\textsc{i}}$) grid of the simulation volume. Starting at scales comparable to the mean free path of ionizing photons in an ionized IGM\footnote{We assume this to be 30~Mpc \citep{SC2010,McQ2011} and to be homogeneous and independent of redshift.} at $z \sim 6$ and incrementing toward smaller scales, voxels with spatial position and redshift $(\vect{x}, z)$ are flagged as being ionized if, in a sphere of radius $R$, the integrated number of ionizing photons is greater than the number of hydrogen and neutral helium atoms plus the mean number of recombinations. In our prescription this ionization condition can be written in terms of an ionization efficiency factor, $\xi$, and the collapsed fraction\footnote{Note that we define the collapsed fraction in a non-standard way using stellar mass since it is directly modelled by \Meraxes. The collapsed fraction is usually defined as the fraction of mass bound in haloes (e.g. $M_{\rm vir}/M_{\rm tot}$) above some minimum. The ionization condition is then formed assuming a baryon fraction of matter, and that a fraction, $f_*$, of this galactic gas is in the form of stars. These fractions are parametrized within the ionization efficiency factor, $\xi$.} of mass, $f_{\rm coll}$, in the form of stars within $R$:
\begin{eqnarray}
\xi f_{\rm coll}(\vect{x}, z, R) \geq 1,
\label{Eqn:ionization_condition}
\end{eqnarray}
where
\begin{eqnarray}
f_{\rm coll}(\vect{x}, z, R) \equiv\frac{M_*(\vect{x}, z, R)}{M_{\rm tot}(\vect{x}, z, R)}.
\label{Eqn:f_coll}
\end{eqnarray}
We set the unresolved, sub-cell neutral hydrogen fraction corresponding to the last smoothing scale in our simulations to $\xi f_{\rm coll}(\vect{x}, z, R_{\rm voxel})$, where $R_{\rm voxel}$ is of the order of the grid resolution of \tocf.

The ionization efficiency factor depends on the baryon fraction, $f_{\rm b}$, mean number of ionizing photons per stellar nucleon, $\bar{N}_\gamma$, and their escape fraction, $f_{\rm esc}$, through
\begin{eqnarray}
\label{eq:ionization_efficiency_factor}
\xi = 6214 \left(\frac{0.157}{f_{\rm b}}\right) \left(\frac{\bar{N}_\gamma}{4000}\right) \left(\frac{f_{\rm esc}}{0.2}\right) \left(\frac{0.82}{1-3Y_{\rm He}/4}\right).
\end{eqnarray}
We set the baryon fraction to its universal value, $f_{\rm b} = \Omega_{\rm b}/\Omega_{\rm m}$. The $(1-3Y_{\rm He}/4)$ factor, where $Y_{\rm He} = 0.24$ is the mass fraction of helium, accounts for helium in the ionization budget. Both $f_{\rm b}$ and $Y_{\rm He}$ are well constrained, while $\bar{N}_\gamma$ is set by an assumed initial mass function\footnote{We use a Salpeter initial mass function for which $\bar{N}_\gamma = 4000$ \citep{BL2001}.} of stars. The value of $f_{\rm esc}$ is less well constrained for galaxies at high redshift (see Section~\ref{Escape fraction} for further discussion). We account for recombinations by commencing the reionization excursion-set filtering at a scale comparable to the mean free path of ionizing photons in an ionized IGM (rather than the simulation box size). This effectively acts as a `global reionization horizon' \citep[see][for a discussion of this in terms of the framework of \tocf]{GREIG2015} and slows reionization during its late phase as seen in the evolution of the globally-averaged neutral fractions in Figure~\ref{global_xH}. Alternatively, it is possible to account for homogeneous recombinations by appropriately including a factor of $(1 + \bar{N}_{\rm rec})$ in Equation~\ref{eq:ionization_efficiency_factor}, where $\bar{N}_{\rm rec}$ is the globally-averaged number of recombinations. Homogeneous recombinations are, however, degenerate with the ionizing efficiency. The effect of inhomogeneous recombinations has been investigated by a number of authors \citep[see, e.g.,][]{CHR2009,SM2014,WMPS2015} and is left for future work.

\subsubsection{Inhomogeneous UV background} 
\label{Inhomogeneous UVB}

In order to simulate the effect of feedback from a UV background on star formation, the \tocf~algorithm calculates a grid of the local average UVB intensity within a given ionized region, $\langle J_{21} \rangle_{\rm H\textsc{ii}}$. Following \cite{SM2013a}, this is given by
\begin{eqnarray}
\label{eq:J21}
\langle J_{21} \rangle_{\rm H\textsc{ii}}(\vect{x}, z, R) = \frac{(1 + z)^2}{4\pi} \lambda_{\rm mfp} h_{\rm P} \alpha f_{\rm bias} \epsilon(\vect{x}, z, R),
\end{eqnarray}
where $\lambda_{\rm mfp}$ is the comoving mean free path of ionizing photons (assumed within the excursion-set algorithm to be equal to the filtering radius, $R$), $h_{\rm P}$ is the Planck constant, $\alpha$ is the spectral index of the UVB, $f_{\rm bias}$ is an ionizing emissivity bias factor and $\epsilon$ is the ionizing emissivity. We model the background spectrum using
\begin{eqnarray}
J(\nu) = J_{21} \left( \nu/\nu_{\rm H} \right)^{-\alpha} \times 10^{-21}\,{\rm erg\,s}^{-1}\,{\rm Hz}^{-1}\,{\rm pcm}^{-2}\,{\rm sr}^{-1},
\end{eqnarray}
where  $\nu_{\rm H} = 3.2872 \times 10^{15}$~Hz is the Lyman-limit frequency, $\alpha = 5$ \citep{TW1996} and pcm denotes proper centimetres. Our choice of spectral index leads to a steep spectrum corresponding to a stellar-driven UVB, effectively ignoring the contribution from harder spectral sources such as quasars. As such, the reionization models presented in this work have been specifically chosen to be galaxy-driven and our results should be interpreted in this context\footnote{While it is sensible to expect a contribution to reionization from quasars, the resulting harder spectrum would still lead to a bi-modality in the ionization state distribution (viz. fully neutral or completely ionized voxels with relatively few that are partially ionized) and inside-out reionization seen in this and other recent work. We are currently working on investigating the effect of including active galactic nuclei (Qin et al., in preparation) and X-rays in our model.}. Furthermore, the form of the UVB at high redshift is still unclear. In fact, recent metal-line observations have been shown to be inconsistent with a hard UVB at $z \sim 6$ \citep{FINLATOR2016}. Regardless, it is interesting to ask what the impact of a harder spectrum on reionization feedback would be. We expect using a much harder spectrum would cause feedback to be more effective \citep[see, e.g.,][]{TW1996}, hence changing our results. However, in light of the dominance of supernova feedback in our simulations, we would not expect a significantly different result. Quantitatively, this would be difficult to estimate without repeating the work of \cite{SM2013a} using a harder spectrum. We leave this investigation to future work. The bias factor in Equation~\ref{eq:J21} accounts for the higher than average ionizing emissivity at halo locations due to their clustering. We use a value of $f_{\rm bias} = 2$ based on discussion in \cite{MD2008}. The ionizing emissivity is the number of ionizing photons emitted into the IGM per unit time, per unit comoving volume. This is calculated for each voxel using the grid of gross stellar mass, $M_{*, \rm gross}$, (the integrated mass of stars formed up to redshift $z$) through
\begin{eqnarray}
\epsilon (\vect{x}, z, R) = f_{\rm esc} \bar{N}_\gamma \frac{M_{*, \rm gross}(\vect{x}, z, R)}{\bar{m}_{\rm b}}  \frac{1}{t_{\rm H}(z)} \frac{1}{V_R},
\end{eqnarray}
where the gross stellar mass has been averaged over a sphere of radius $R$ with volume $V_R$, $t_{\rm H}(z)$ is the Hubble time which acts as the average star-formation time-scale, and $\bar{m}_{\rm b}$ is the mean mass of each baryon (approximately equal to the mass of the proton, $m_{\rm p}$).

\subsubsection{Feedback}
\label{Feedback}

A UV background can lead to negative feedback on cosmic reionization in various ways. Its presence reduces baryonic infall by way of heating the IGM, therefore affecting its gas cooling properties. It can also photo-evaporate gas from shallow potential wells surrounding small galaxies. Both of these mechanisms lead to the quenching of star formation. We parametrize this effect in \Meraxes~using a spatially- and temporally-dependent baryon fraction modifier, $0 \le f_{\rm mod} \le 1$, which acts to reduce the mass fraction of baryons contained in freshly accreted matter from its universal value, $f_{\rm b}$, to $f_{\rm mod} f_{\rm b}$. Following the work of \cite{SM2013b}, we calculate $f_{\rm mod}$ using
\begin{eqnarray}
\label{eq:fmod}
f_{\rm mod}(\vect{x}, z) = 2^{-M_{\rm filt}/M_{\rm vir}},
\end{eqnarray}
where $M_{\rm vir}$ is the halo mass and $M_{\rm filt}$ is a `filtering' mass, defined to be the total halo mass at which the baryon fraction is half the universal value. Together with the snapshot redshift, $z$, and the redshift at which the voxel was first ionized, $z_{\rm ion}$, the local average UVB intensity within a given ionized region is used to calculate the filtering mass:
\begin{eqnarray}
M_{\rm filt} = M_0 J_{21}^a \left( \frac{1 + z}{10} \right)^b \left[ 1 - \left(\frac{1 + z}{1 + z_{\rm ion}}\right)^c \right]^d,
\end{eqnarray}
where $(M_0, a, b, c, d) = (2.8 \times 10^9\,M_\odot, 0.17, -2.1, 2.0, 2.5)$. These free parameters were fit by \cite{SM2013b} to one-dimensional collapse simulations.

\subsubsection{Escape fraction}
\label{Escape fraction}

Not all of the ionizing photons produced by stars within galaxies manage to escape their hosts so that they may contribute to the ionization of the IGM. The presence of dust and neutral gas within and around galaxies reduces the ionizing efficiency of these sources. The ionizing efficiency of very high-redshift galaxies is poorly constrained by observations of galaxy UV luminosity functions and our knowledge of the escape fraction of ionizing photons and the abundance of faint galaxies is poor. Furthermore, simulations show a strong mass dependency and anisotropy of the escape fraction \citep[e.g.][]{PAAR2015}, both of which can strongly impact the topology of reionization. This lack of observational constraint and expected complexity allows for a wide variety of plausible reionization scenarios \citep[examples of recent work in this area include studies by, e.g.,][]{KFG2012,KIM2013b,PAAR2015}. As shown in Paper-III, while we are able to match high-redshift galaxy stellar mass functions and satisfy constraints imposed by CMB measurements using an escape fraction which does not evolve with redshift, we are unable to also match the normalisation and slope of the observed ionizing emissivity at $z \le 6$ without using a redshift-dependent escape fraction. Furthermore, Lyman-$\alpha$ emitter studies suggest that reionization ends at $z \sim 6$ \citep[see, e.g.,][]{DIJKSTRA2014,CPHB2015}. An evolving escape fraction prolongs reionization and delays its completion, hence simultaneously satisfying these constraints. Similarly, inhomogeneous recombinations \citep[\`a la][]{SM2014} are expected to prolong the late stages of reionization. However, for the purpose of comparing models presented in this work, we have kept our prescription for the escape fraction as simple as possible by assuming it to be spatially and temporally invariant.

\begin{table*}\centering
\ra{1.2}
\begin{tabular}{p{0.4cm} p{1.9cm} p{7cm} c c c c}
\toprule
& Model & Description & $f_{\rm esc}$ & $\alpha_{\rm {SF}}$ & Relative av. & Max. frac. diff.\\
&  &  &  & & bubble size & in dim. power\\
\midrule
\midrule
\raisebox{1mm}{\includegraphics[width=0.5cm]{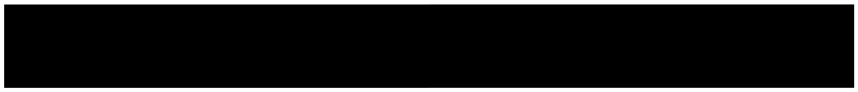}} & F (Fiducial) & Feedback from a UV background incorporated as per Section~\ref{Inhomogeneous UVB}. Supernova feedback incorporated as per Section~\ref{Semi-analytic modelling}. & 0.2 & 0.03 & - & -\\
\midrule
\raisebox{1mm}{\includegraphics[width=0.5cm]{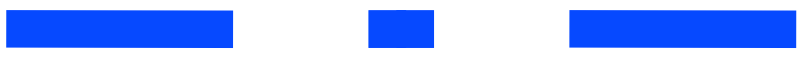}} & NoSNeFB & Recalibrated fiducial model without feedback from supernov\ae. & 0.239 & 0.00106 & 19\% $\downarrow$ & 17\%\\
\midrule
\raisebox{1mm}{\includegraphics[width=0.5cm]{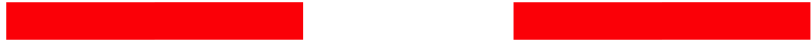}} & NoFB & Recalibrated fiducial model without feedback from reionization or supernov\ae. & 0.2328 & 0.00106 & 18\% $\downarrow$ & 15\%\\
\midrule
\raisebox{1mm}{\includegraphics[width=0.5cm]{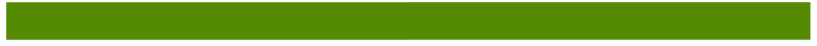}} & CSHR & Constant stellar-to-halo mass ratio set as $M_* / M^{\rm FoF}_{\rm vir} = 0.055$. & 0.01547 & 0.03 & 22\% $\downarrow$ & 18\%\\
\midrule
\raisebox{1mm}{\includegraphics[width=0.5cm]{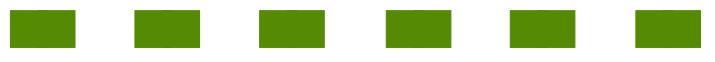}} & CSHR.Mcut.9 & Constant stellar-to-halo mass ratio model only including galaxies with $M^{\rm FoF}_{\rm vir} \geq 10^{9}$~M$_\odot$. & 0.031 & 0.03 & 17\% $\downarrow$ & 15\%\\
\midrule
\raisebox{1mm}{\includegraphics[width=0.5cm]{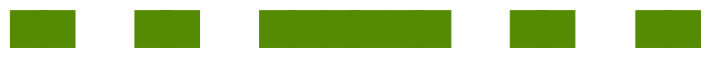}} & CSHR.Mcut.10 & Constant stellar-to-halo mass ratio model only including galaxies with $M^{\rm FoF}_{\rm vir} \geq 10^{10}$~M$_\odot$. & 0.1302 & 0.03 & 19\% $\uparrow$ & 69\%\\
\bottomrule
\end{tabular}
\caption{Simulation model and results summary. Note that we tune the escape fraction in each simulation to give a neutral hydrogen volume fraction of $\approx 0.68$ at $z = 8.4$ to facilitate comparison between models. Also listed is the resulting relative average `bubble' size with respect to the fiducial model ($\uparrow$ and $\downarrow$ indicating an increase and decrease, respectively), and the maximum fractional difference in dimensional 21-cm power with respect to the fiducial model in the range $k = 0.2$--1~Mpc$^{-1}$. Both sets of results are for comparisons made at $\bar{x}_{\rm H\textsc{i}} \approx 0.68$. See Sections~\ref{Reionization morphology} and \ref{21cm power spectra} for a more detailed discussion.}
\label{sim_summary_table}
\end{table*}

\subsection{Model simulations}
\label{Simulations}

In this section we describe the set of models used in this work. A summary is listed in Table~\ref{sim_summary_table}. Models that ignore the effects of supernova feeback have been recalibrated so as to provide the same total $z = 5$ stellar mass density as the fiducial model (see discussion in Paper-III). This has been achieved by using a lower star-formation efficiency parameter, $\alpha_{\rm {SF}}$. In order to facilitate direct comparison between our models, we `tune' the ionization efficiency factor of each by varying the escape fraction of ionizing photons so as to match the globally-averaged neutral fraction of our fiducial model at $z \approx 8.4$ ($\bar{x}_{\rm H\textsc{i}} \approx 0.68$, see Table~\ref{sim_summary_table}). This results in the models having very similar reionization histories (see Figure~\ref{global_xH}). Therefore, with each model having a similar global neutral fraction and the same underlying density field at each snapshot/redshift, we can sensibly compare models by way of their 21-cm power spectra (which are sensitive to both neutral fraction \emph{and} density). We choose to match at $\bar{x}_{\rm H\textsc{i}} \approx 0.7$ as our cosmic volumes at this stage of reionization provide a fair sample of different \HIItext~region sizes and forms (as seen in the ionization maps in Figure~\ref{xHI_maps}).

\noindent \textbf{Fiducial model (F):} We couple both the spatial and temporal evolution of the reionization structure and ionizing field to the growth of the source galaxy population through the incorporation of feedback by an inhomogeneous UVB. Supernova feedback is included and a spatially-homogeneous, redshift-independent escape fraction of ionizing photons of $f_{\rm esc} = 0.2$ is used. This model is identical to the fiducial model in Paper-III and is discussed at length in that paper and Paper-IV.

\noindent \textbf{No SN feedback (NoSNeFB):} This is a recalibrated version of the fiducial model which ignores the effect of feedback from supernov\ae. A constant $f_{\rm esc} = 0.239$ is used.

\noindent \textbf{No feedback (NoFB):} This is a recalibrated version of the fiducial model which ignores the effect of feedback from both reionization and supernov\ae. A constant $f_{\rm esc} = 0.2328$ is used.

\noindent \textbf{Constant stellar-to-halo mass ratio (CSHR):} This model ignores the galaxy properties calculated by \Meraxes~and is therefore decoupled from the effects of any form of feedback. Stellar mass is calculated assuming a constant ratio between stellar mass and the virial mass of each Friends-of-Friends (FoF) halo group using $M_* / M_{\rm vir} = 0.055$. This is based on the value we find in the high-mass regime of a simulation with no feedback from either reionization or supernov\ae. This model (along with the following mass-cut models) is included so as to facilitate comparison of our fiducial model simulation with previously published semi-numerical simulations and models that do not include realistic galaxy formation and feedback effects. A constant $f_{\rm esc} = 0.01547$ is used.

\noindent \textbf{Constant stellar-to-halo mass ratio with $10^9$~M$_\odot$ mass cut (CSHR.Mcut.9):} This is the same as the CSHR model but only includes galaxies whose FoF virial mass is $M^{\rm FoF}_{\rm vir} \geq 10^{9}$~M$_\odot$. A constant $f_{\rm esc} = 0.0312$ is used.

\noindent \textbf{Constant stellar-to-halo mass ratio with $10^{10}$~M$_\odot$ mass cut (CSHR.Mcut.10):} This is the same as the CSHR model but only includes galaxies whose FoF virial mass is $M^{\rm FoF}_{\rm vir} \geq 10^{10}$~M$_\odot$. A constant $f_{\rm esc} = 0.1302$ is used.

\section{Results}
\label{Results}

\subsection{Reionization history}
\label{Reionization history}

\begin{figure}
\includegraphics[width= 8.5cm]{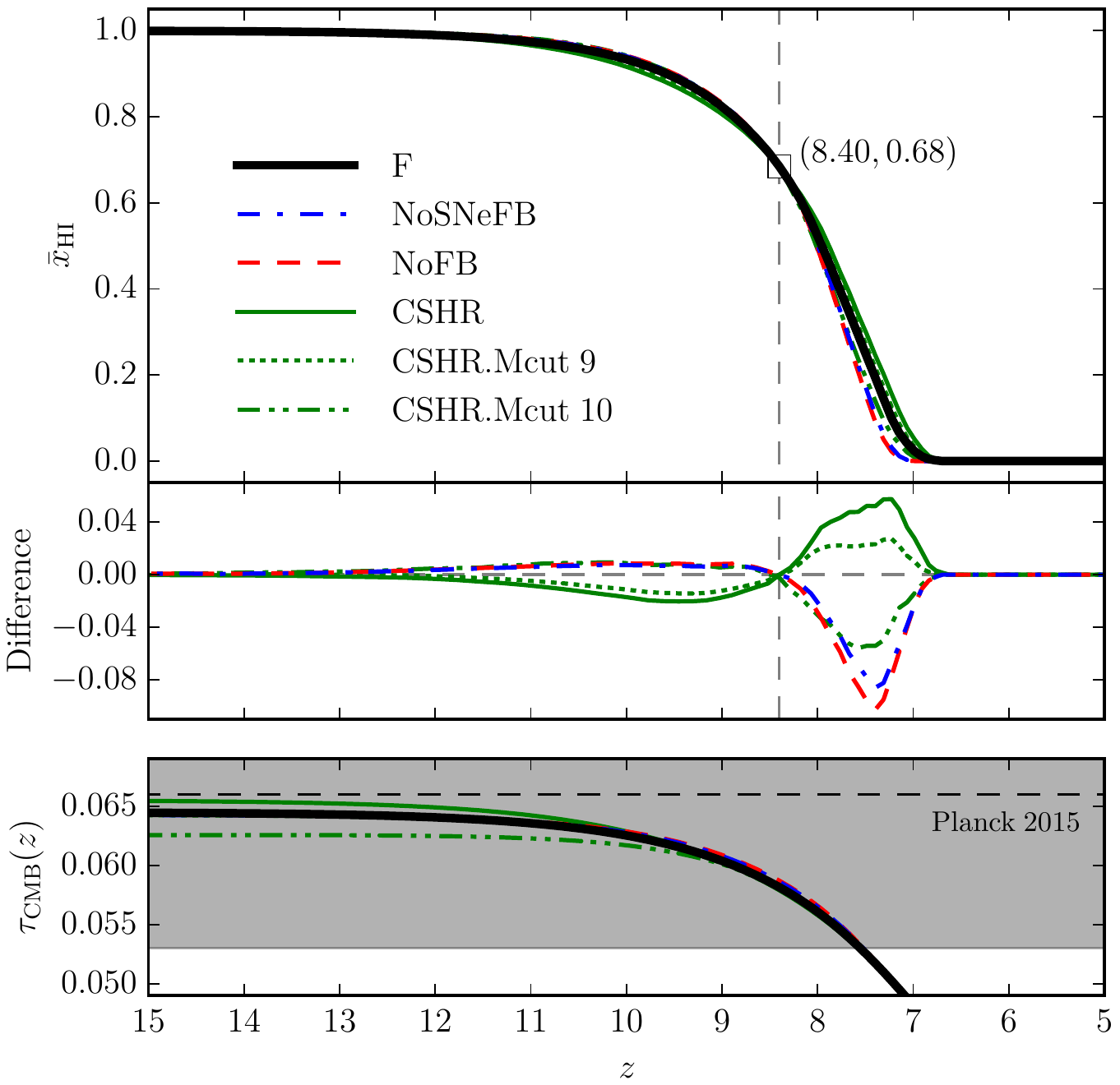}
\caption{Top panel: Globally-averaged neutral fraction, $\bar{x}_{\rm H\textsc{i}}$, as a function of redshift for our models. Middle panel: Difference between the globally-averaged neutral fraction of each model with respect to the fiducial model as a function of redshift. The vertical dashed line shows the redshift at which the global neutral fractions of the models have been matched. Bottom panel: Cumulative Thomson scattering optical depth seen by CMB photons as a function of integrated redshift for our models (due to their similarity, the no-feedback models' lines lie behind the fiducial model line). The horizontal dashed line and surrounding shaded region mark the latest \textit{Planck} observations and $\pm1\sigma$ uncertainty.}
\label{global_xH}
\end{figure}

Figure~\ref{global_xH} shows the evolution of the globally-averaged neutral fraction, $\bar{x}_{\rm H\textsc{i}}$, of each model (top panel) and their difference with respect to the fiducial model (middle panel) as a function of redshift. Reionization for our fiducial model occurs over a period of 411~Myr (period between the simulation volume being 1--99~per~cent ionized) and is 99~per~cent ionized at $z \approx 6.9$. The data point in Figure~\ref{global_xH} indicates the neutral fraction at which the models have been matched. This figure provides a comparison between the duration and rates of reionization for each model. Each model follows a similar reionization profile, the greatest difference (either time-averaged or for any single snapshot) being for the NoFB model. The CSHR model reionizes least rapidly, owing to the more complete distribution of ionizing source masses for this model (i.e. more stellar material in low-mass haloes than the other models; see the discussion in Section~\ref{Reionization morphology}).

The bottom panel in Figure~\ref{global_xH} shows the cumulative CMB optical depth to Thomson scattering as a function of integrated redshift for our reionization model simulations (the no-feedback models' results are very similar to those of the fiducial model and lie behind its line in this figure). These results have been calculated in the same manner as in Paper-III. The horizontal dashed line and surrounding shaded region mark the latest \textit{Planck} observations and $\pm1\sigma$ uncertainty \citep{PLANCK2015}. The total CMB optical depth to electron scattering (high-redshift limit) of our fiducial model falls within this observational constraint. While the other models have had their reionization histories matched to the fiducial model, small differences in their reionization profiles result in a small variation in their optical depths. All results, however, fall well within the constraints provided.

\begin{figure*}
\includegraphics[width= 17.6cm]{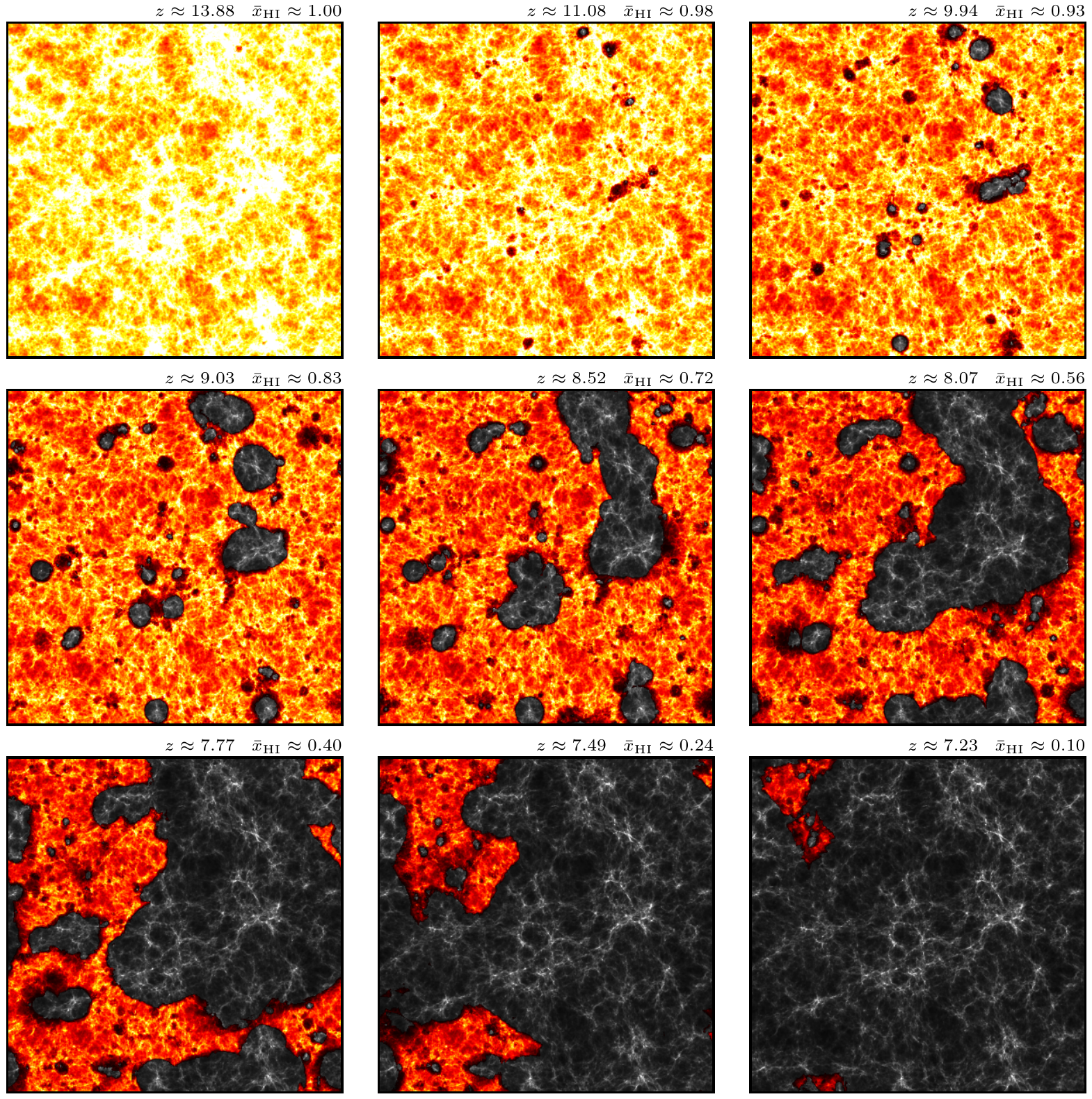}
\caption{Neutral gas density slices for our fiducial model at selected redshifts. The underlying dark matter density contrast is shown in fully ionized regions (`bubbles'). The spatial correlation between the higher-density cosmic web and regions of ionized gas is clear. The redshift and globally-averaged neutral fraction is shown above each panel. All slices are 100~Mpc on a side and 4~Mpc deep.}
\label{Tb_maps}
\end{figure*}

The spatial progression of reionization can be seen in Figure~\ref{Tb_maps} which shows slices of the neutral gas density for our fiducial model at selected redshifts. The underlying dark matter density contrast is shown in fully ionized regions (`bubbles').

\subsection{Reionization morphology}
\label{Reionization morphology}

\begin{figure*}
\includegraphics[width= 17cm]{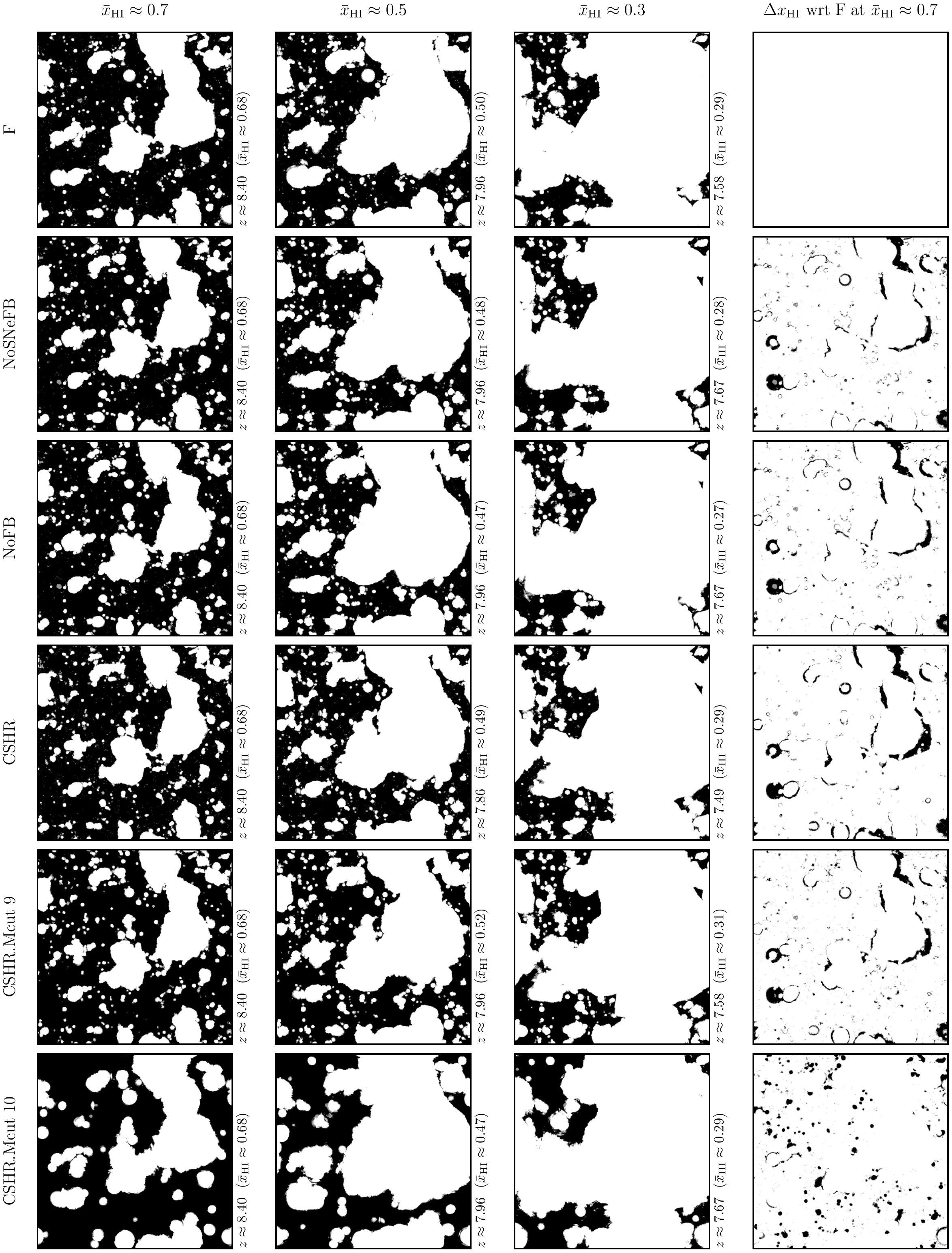}
\caption{Ionization field of our models (rows) at three different global neutral fractions ($\bar{x}_{\rm H\textsc{i}} \approx 0.7$, 0.5 and 0.3, left to right respectively). White regions are ionized and black regions are neutral. At $\bar{x}_{\rm H\textsc{i}} \approx 0.7$, the NoSNeFB, NoFB, CSHR, CSHR.Mcut.9 and  CSHR.Mcut.10 models result in regions that are $\approx 19$~per~cent smaller, 18~per~cent smaller, 22~per~cent smaller, 17~per~cent smaller and 19~per~cent larger than those in the fiducial model, respectively. The far-right panel shows the difference between the ionization field of each model with respect to that of the fiducial model. All panels are 100~Mpc on a side and 0.4~Mpc deep.}
\label{xHI_maps}
\end{figure*}

In this section we focus on the morphology of reionization, and compare the effect of different ionizing source populations and feedback mechanisms. The corresponding results for 21-cm power-spectra statistics are discussed in Section~\ref{21cm power spectra}.

Reionization on intergalactic scales is expected to proceed in an `inside-out' manner whereby more-over-dense regions (e.g. sites of the first-formed galaxies) reionize before less-over-dense regions \citep[see, e.g.,][]{ILIEV2006,MLZD2007,MFC2011,BAUERetal2015}. The spatial correlation between the higher-density cosmic web and regions of ionized hydrogen is clear in the neutral gas density slices of our fiducial model shown in Figure~\ref{Tb_maps}.

The effect of feedback and different ionizing source populations can be seen in Figure~\ref{xHI_maps}, which shows slices through the ionization field of our models at three different global neutral fractions ($\bar{x}_{\rm H\textsc{i}} \approx 0.7$, 0.5 and 0.3, left to right respectively). In this figure, white regions are ionized and black regions are neutral. Also shown are the differences between the ionization fields of each model with respect to that of the fiducial model (far-right panel).

\begin{figure}
\includegraphics[width= 8.5cm]{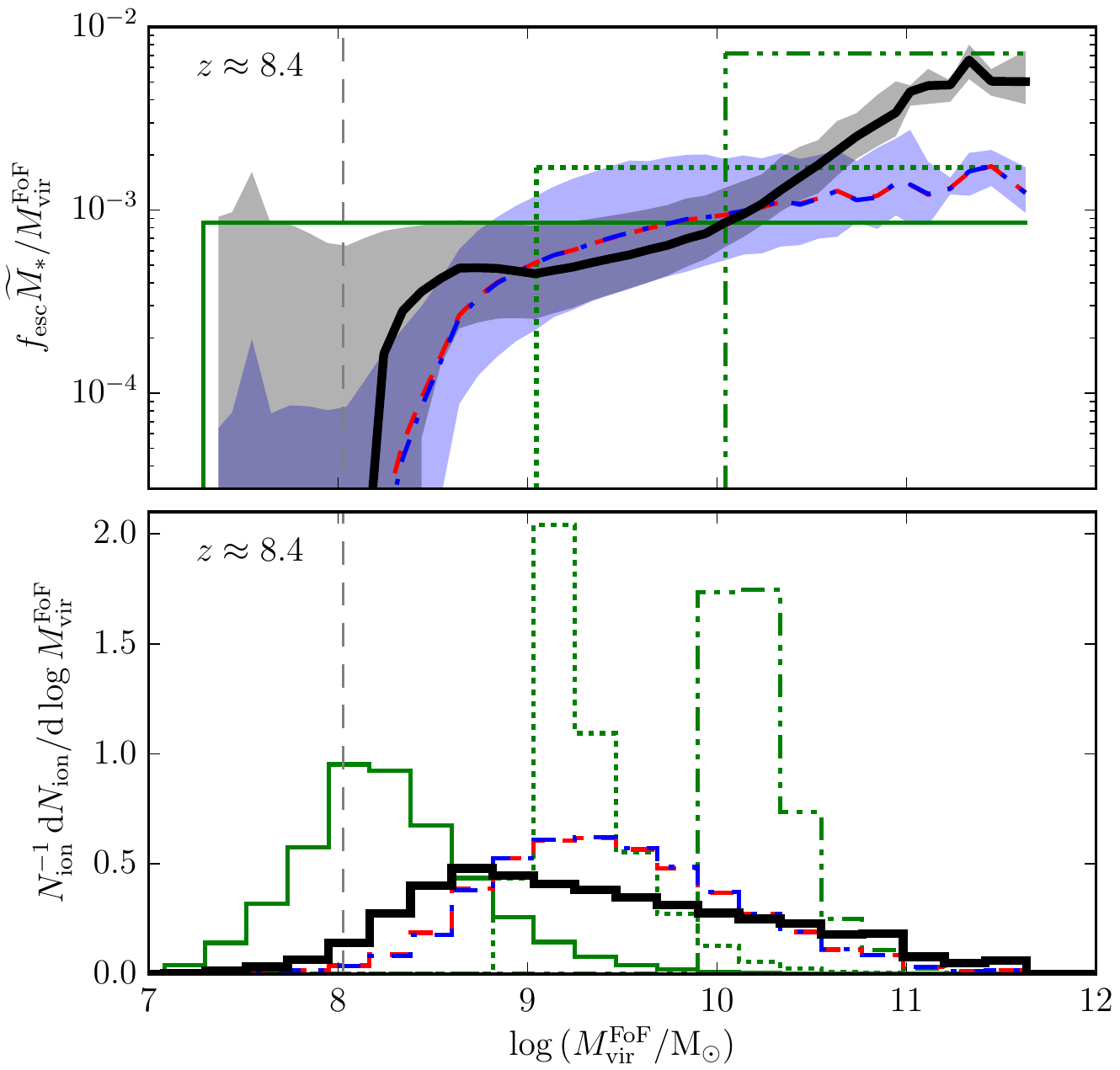}
\caption{Top panel: Median fraction of mass in the form of stars at $z \approx 8.4$ as a function of FoF virial mass, weighted by the escape fraction of each model. This is used as a proxy for the average ionizing luminosity of sources in our models. Shaded regions indicate the 68~per~cent confidence range (not shown for the NoFB model due to its similarity to the NoSNeFB model). Bottom panel: Probability distributions of the integrated ionizing photon contribution up to $z \approx 8.4$ as a function of FoF virial mass for each model. The vertical dashed lines show the atomic cooling mass threshold at this redshift, below which no stars form. As a result of mass stripping upon merger infall and the presence of galaxies that have formed at earlier times (when the atomic cooling mass threshold was lower), a relatively small number of galaxies reside in haloes below this threshold.}
\label{FESCweightedSHMR}
\end{figure}

In order to help explain differences and similarities between the ionization fields of our models, we refer to the plots in Figure~\ref{FESCweightedSHMR} which show the following ionization properties of their source populations. In the top panel we have calculated the median fraction of mass in the form of stars at $z \approx 8.4$ as a function of FoF virial mass, weighted by the escape fraction of each model, i.e. $f_{\rm esc} \widetilde{M}_* / M^{\rm FoF}_{\rm vir}$. This quantity acts as a proxy for the average ionizing luminosity of sources as a function of their host halo virial mass\footnote{On first sight, the top panel of Figure~\ref{FESCweightedSHMR} appears to contradict the star formation suppressing effects of feedback. However, misinterpretation is avoided by keeping in mind that the no feedback models have been recalibrated using a lower star-formation efficiency factor. The product of $\alpha_{\rm SF}f_{\rm esc}$ (rather than $f_{\rm esc}$ alone) for each model is consistent with the effects of feedback.} but includes no information about the size of their populations. In the bottom panel we show the probability distributions of the integrated ionizing photon contribution up to $z \approx 8.4$ as a function of FoF virial mass for each model. As these histograms have been weighted by the gross stellar mass in each halo mass bin, the distributions include the effect of source population.

The following features can be observed in the ionization fields shown in Figure~\ref{xHI_maps}:
\begin{itemize}
\item \textit{The ionization field for the NoSNeFB model has a slightly larger population of small (1--3~Mpc) ionized regions and its overlapping regions tend to be smaller than those of the fiducial model.} As evident from the top panel of Figure~\ref{FESCweightedSHMR}, including the star-formation suppressing effect of SNe feedback \emph{and} tuning $f_{\rm esc}$ so as to match the models' reionization states effectively lowers the specific luminosity of sources hosted by medium-mass ($10^9 \lsim~M_{\rm vir} \lsim~10^{10}$~M$_\odot$) haloes, but increases it for the very lowest- and highest-mass sources. Due to their relative populations, these medium-mass sources are the dominant drivers of reionization in the NoSNeFB model up to this period of reionization (as can be seen in the bottom panel of Figure~\ref{FESCweightedSHMR}). As these haloes are less biased than more-massive haloes, they tend to populate the less-dense regions of the simulation volume away from the clustered sources central to the overlapping ionized regions. The relatively small ionizing contribution from sources hosted by the  very lowest- and highest-mass haloes play little to no role toward differences in morphology. Therefore, \textit{without} the feedback effects of supernov\ae, more small isolated ionized regions will exist and the overlapping regions will be smaller at fixed ionized fraction. This is in agreement with previous work investigating SNe feedback \citep[see, e.g.,][]{KIM2013a}, although the effects are significantly weaker in our results. We attribute this difference to \emph{Tiamat} having an order of magnitude higher temporal resolution than the simulation used by \cite{KIM2013a}. This resolution is required to resolve the dynamical time at $z > 6$.
\item \textit{The ionization field for the NoFB model is almost identical to that of the NoSNeFB model.} This is due to the dominance of supernova feedback over reionization feedback in suppressing star formation as discussed in Paper-III. Therefore, the same general argument holds here in explaining the difference in the ionization fields as for the NoSNeFB model above.
\item \textit{The ionization field for the CSHR model has a larger population of small ionized regions, including very small ($<$\,1~Mpc) regions (although these are somewhat difficult to see in Figure~\ref{xHI_maps}), and its overlapping regions tend to be smaller than those of the fiducial model.} Galaxies with low-mass FoF hosts in the \Meraxes~models contain, on average, less stellar mass than that expected using a constant stellar-to-halo mass ratio prescription. This is due to both the galaxy-formation modelling, and star-formation suppressing effects of reionization and (predominantly) supernova feedback (as discussed in Paper-III). As evident in the bottom panel of Figure~\ref{FESCweightedSHMR}, sources hosted by these low-mass haloes in the CSHR model dominate reionization\footnote{The drop-off at $M_{\rm vir} < 10^8$~M$_\odot$ is solely due to the halo mass resolution limit of \emph{Tiamat} (see the halo mass functions in Figure~A1 in the appendix of Paper-III.}. For the same reasons outlined in the argument for the no-feedback model results above, this leads to a larger population of small isolated ionized regions, but with a greater number of very small regions. Despite the relatively large shift in the host halo mass scale of sources which dominate reionization between the \Meraxes-based models and the CSHR model, the resulting ionization fields are remarkably similar. This was the motivation behind the CSHR mass-cut models (see Section~\ref{Discussion} for a discussion on the pros and cons of these two prescriptions).
\item \textit{The ionization field of the CSHR.Mcut.9 model bears a striking resemblence to those of the no-feedback and CSHR models, but contains fewer very small isolated ionized regions.} Again, the same general argument holds here as for the no-feedback and CSHR models. The mass cut imposed on this model removes the ionizing contribution of sources hosted by the lowest-mass haloes present in the fully-populated CSHR model but, as evident in the bottom panel of Figure~\ref{FESCweightedSHMR}, leaves reionization to be dominated by sources hosted by haloes in the same medium-mass range as the no-feedback models (hence their similarity). The importance of this result is discussed further in Section~\ref{Discussion}.
\item \textit{Ionized regions in the CSHR.Mcut.10 model are larger, more clustered and more spherical than those of the other models.} On average, ionizing sources of the CSHR.Mcut.10 model are more luminous than those in all of our other models (as demonstrated in the upper panel of Figure~\ref{FESCweightedSHMR}). Since only high-mass ionizing sources have been included in this model (by way of the mass cut of their hosts) and these sources are more biased (again, by way of their hosts), they tend to cluster within the densest environments, forming large overlapping ionized regions. Without the presence of smaller ionized regions formed by less-luminous (and less-biased) sources, isolated regions tend to appear more spherical.
\end{itemize}

\begin{figure}
\includegraphics[width= 8.5cm]{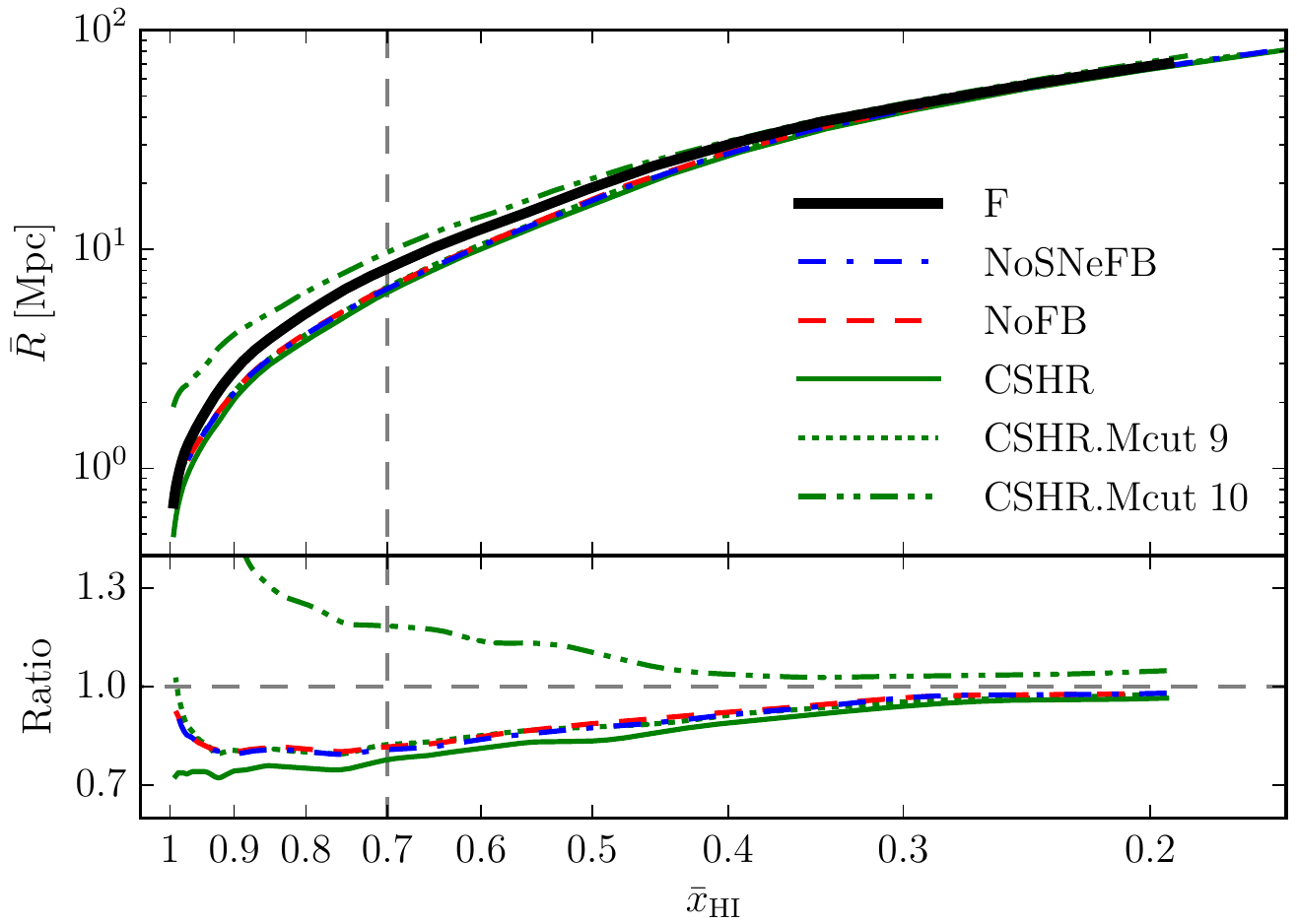}
\caption{Top panel: Mean size of ionized regions, $\bar{R}$, as a function of global neutral fraction, $\bar{x}_{\rm H\textsc{i}}$. Bottom panel: Ratio of each model's mean size of ionized regions to that of the fiducial model (results have been interpolated in $\log \bar{x}_{\rm H\textsc{i}}$ space). The vertical dashed line shows the global neutral fraction at which the models have been matched.}
\label{bubble_stats}
\end{figure}

In order to quantify these differences in the real-space morphology of reionization we calculate the ionized region (or `bubble') size distributions for each snapshot using the Monte Carlo method described in \cite{MF2007}. In this method, an ionized voxel is randomly selected and its distance from an ionization phase transition (demarked by a step to a voxel that is not completely ionized) in a randomly chosen direction is recorded. This is repeated $10^7$ times to form a probability distribution function of region size. This methodology provides an approximate measure of the mean free path of ionizing photons inside ionized regions and has been extensively used as a proxy for bubble radius in other work \citep[e.g.][]{FO2005,MLZD2007,MF2007}. The top panel of Figure~\ref{bubble_stats} shows the mean size of ionized regions, $\bar{R}$, calculated using this method, as a function of global neutral fraction for each model. The bottom panel shows the ratio of the resulting mean scale of ionized regions relative to the fiducial model (obtained by interpolating the results over $\bar{x}_{\rm H\textsc{i}}$). This figure quantitatively supports the qualitative results illustrated in Figure~\ref{xHI_maps}.

\begin{figure}
\includegraphics[width= 8.5cm]{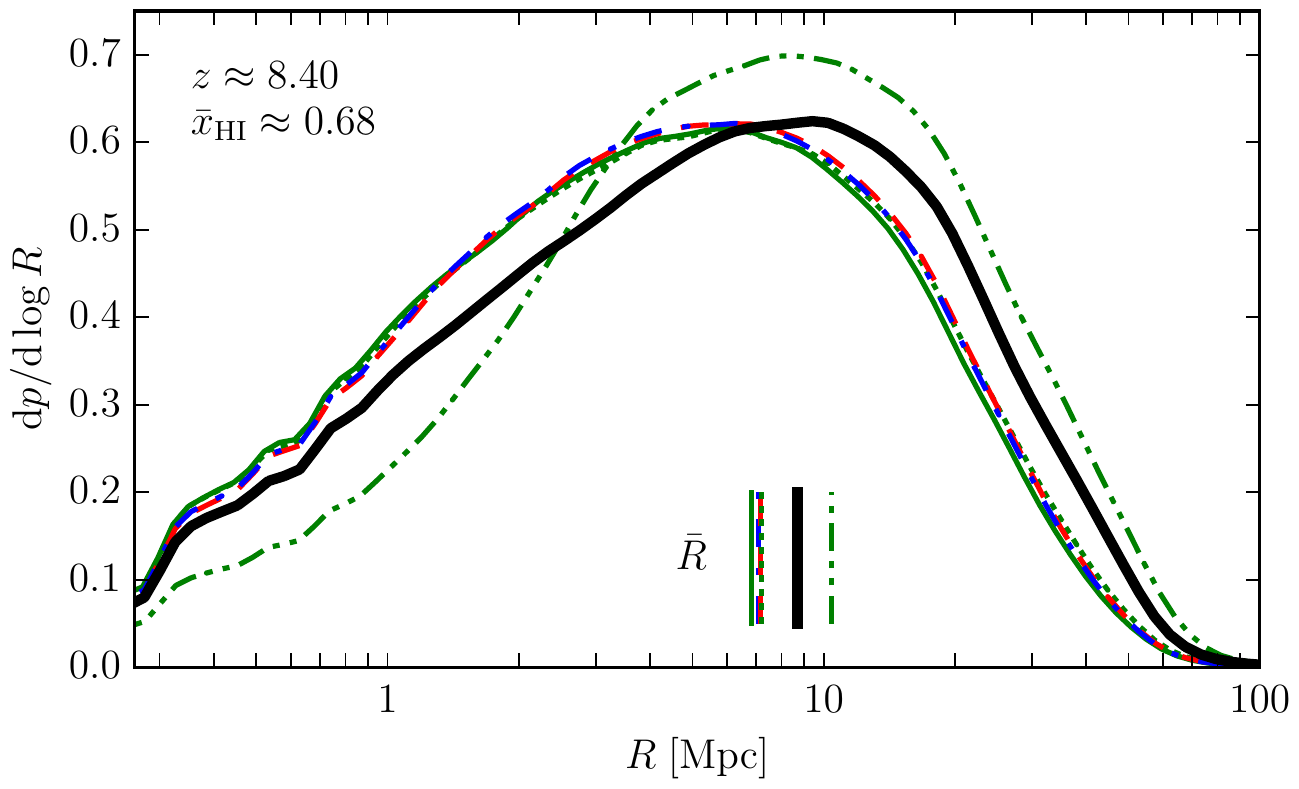}
\caption{Probability distribution of ionized region size for our models at a globally-averaged neutral fraction of $\bar{x}_{\rm H\textsc{i}} \approx 0.68$. Also shown is the corresponding mean of each distribution, $\bar{R}$. The same line styles  as given in Table~\ref{sim_summary_table} have been used.}
\label{bubble_pdfs}
\end{figure}

We show the probability distribution of ionized region size for our models at a globally-averaged neutral fraction of $\bar{x}_{\rm H\textsc{i}} \approx 0.68$ in Figure~\ref{bubble_pdfs}. Also shown is the corresponding mean of each distribution, $\bar{R}$. Relative to the fiducial model, we find the following comparisons of average `bubble' size: i) without SNe feedback, regions are $\approx 19$~per~cent smaller; ii) without SNe or reionization feedback, regions are $\approx 18$~per~cent smaller; iii) using a constant stellar-to-halo mass relationship results in regions that are $\approx 22$~per~cent smaller, iv) using a constant stellar-to-halo mass relationship, but including only sources with haloes of mass greater than $10^9$~M$_\odot$, regions are $\approx 17$~per~cent smaller, and; v) using a constant stellar-to-halo mass relationship but including only sources with haloes of mass greater than $10^{10}$~M$_\odot$ leads to regions that are $\approx 19$~per~cent larger.

\subsection{Cosmic 21cm signal}
\label{Cosmic 21cm signal}

\subsubsection{21-cm differential brightness temperature}
\label{21-cm differential brightness temperature}

We calculate the spatially-dependent 21-cm differential brightness temperature, $\delta T_{\rm b}$, between hydrogen gas and the CMB along the line of sight \citep[for a detailed discussion of the fundamental physics of the 21-cm line, see, e.g.,][]{FOB2006}. For $z \gg 1$, the evolution of $\delta T_{\rm b}$ can be written as
\begin{eqnarray}
\delta T_{\rm b} \approx 27 x_{\rm H\textsc{i}} (1 + \delta) \left( \frac{1 + z}{10} \frac{0.15}{\Omega_{\rm m}h^2} \right)^{1/2} \left( \frac{\Omega_{\rm b}h^2}{0.023} \right)\ {\rm mK},
\label{eq:deltaTb}
\end{eqnarray}
where $\delta = \delta(\vect{x}, z) \equiv \rho(\vect{x}, z)/\bar{\rho}(z) - 1$ is the local dark matter overdensity and $x_{\rm H\textsc{i}} = x_{\rm H\textsc{i}}(\vect{x}, z)$ is the local neutral fraction of the gas at position $\vect{x}$ and redshift $z$. By using this formulation we ignore redshift-space distortions and assume the spin temperature of the cosmic gas, $T_{\rm s}$, to be much higher than the CMB temperature, $T_\gamma$. As a result, this signal is seen in emission. The latter assumption is valid in the post-heating regime where X-ray heating and the Lyman-$\alpha$ background act to decouple the 21-cm transition from the CMB \citep[see, e.g., the discussion by][]{FOB2006}. Caution must be taken, however, if claiming the spin temperature can be neglected during the early stages of reionization since the effect of X-rays on the 21-cm signal depends on the timing of the X-ray heating epoch. As discussed in \cite{MFS2013}, an overlap of these epochs can lead to 10--100 times more power at $k \approx 0.1$~Mpc$^{-1}$ at $\bar{x}_{\rm H\textsc{i}} \gsim~0.9$ than predicted assuming $T_{\rm s} \gg T_\gamma$. At $\bar{x}_{\rm H\textsc{i}} \approx 0.7$, this assumption is found to overpredict 21-cm power by a factor of 1--2. Therefore, while we show results for $z \lsim\ 15$ only in this work, these effects must be kept in mind. They should, however, not affect our model-relative results as we compare the power spectra of our models in terms of their power \emph{ratio} at the single global neutral fraction $\bar{x}_{\rm H\textsc{i}} \approx 0.68$.

\subsubsection{21cm power spectra}
\label{21cm power spectra}

The spherically-averaged dimensionless 21-cm power spectrum is a measure of the relative power of fluctuations in the 21-cm brightness temperature field as a function of scale:
\begin{eqnarray}
\Delta^2_{21}(k, z) = \frac{k^3}{2\pi^2 V} \langle |\hat{\delta}_{21}(\vect{k}, z)|^2 \rangle_k.
\label{eq:21cmPS}
\end{eqnarray}
Here $V$ is the volume of the simulation and $\hat{\delta}_{21}(\vect{k}, z)$ is the Fourier transform of $\delta_{21}(\vect{x}, z)$, calculated using\footnote{Other formulations exist in the literature, including $\delta T_{\rm b}(\vect{x}, z) / \overline{\delta T}_{\rm b}(z)$, and (somewhat confusingly) $\delta T_{\rm b}(\vect{x}, z) / T_0(z)$, where $T_0(z) = 27[(1+z)/10]^{1/2}$~mK (i.e. the brightness temperature of completely neutral gas of average density). The former definition gives the same power-spectral results as for our formulation if the DC mode is not included in spectral binning (i.e. power appearing at $k = 0$, arising from a constant non-zero offset in the $\delta T_{\rm b}$ field).}
\begin{eqnarray}
\delta_{21}(\vect{x}, z) = \frac{\delta T_{\rm b}(\vect{x}, z)}{\overline{\delta T}_{\rm b}(z)} - 1.
\label{eq:delta21}
\end{eqnarray}
A dimensional form of the 21-cm power spectrum (with units of mK$^2$) can be obtained through $\overline{\delta T}_{\rm b}^2 \Delta^2_{21}$, where $\overline{\delta T}_{\rm b}$ is the mean value of the brightness temperature field at redshift $z$. Both can be calculated from observational data but the dimensional power spectrum has the advantage of including the effect of the global ionization state on the overall power. Simulations show that the variance of the brightness temperature field reaches a maximum approximately midway through the reionization process \citep[see, e.g.,][]{LIDZ2008,ICHIKAWA2010,WP2014}. These maxima are also present in the dimensional 21-cm power spectra but not in the dimensionless power spectra. This feature may enable the observational determination of the redshift at which the IGM is approximately half ionized.

\begin{figure}
\includegraphics[width= 8.5cm]{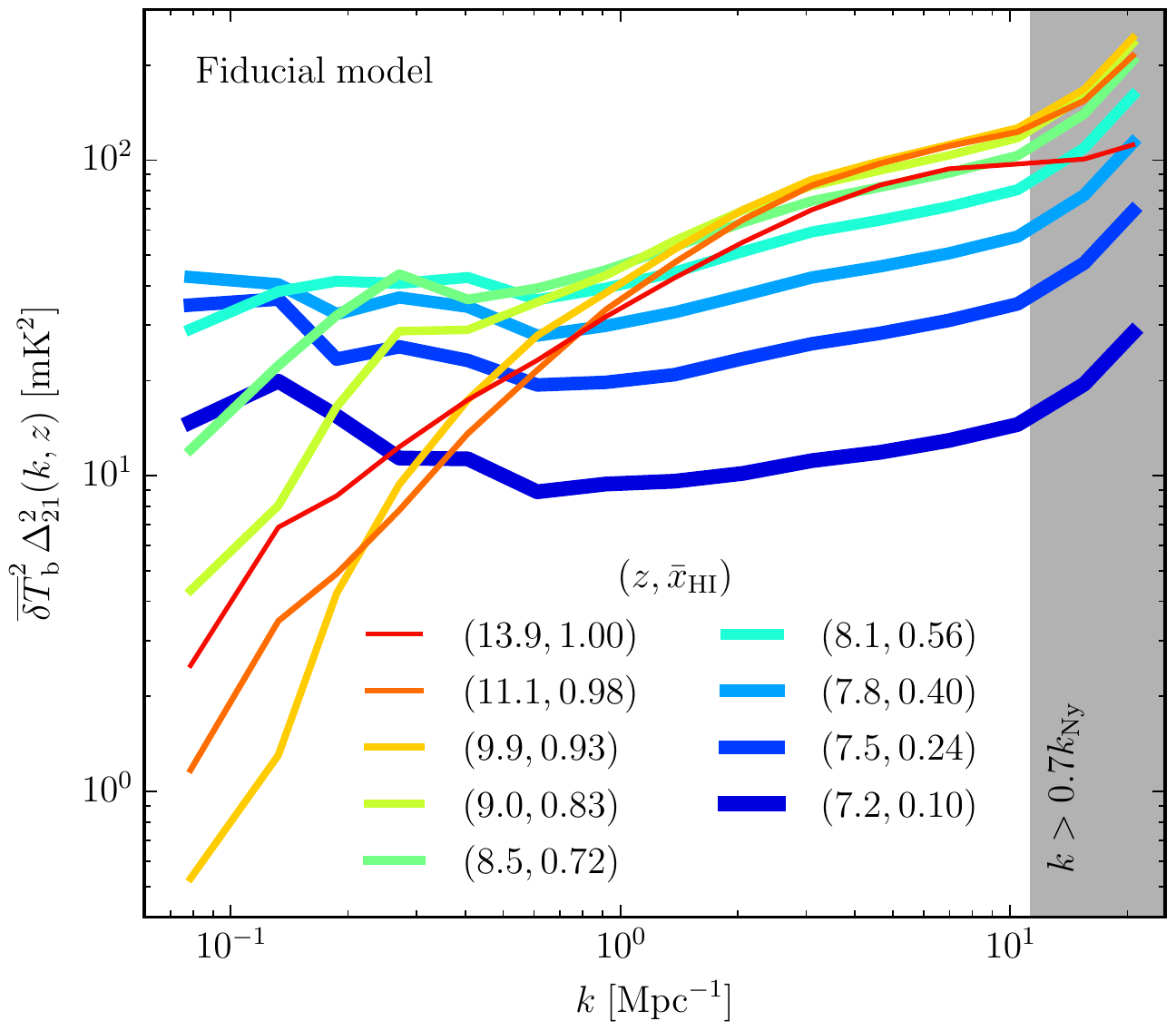}
\caption{Evolution of the dimensional 21-cm power spectra for our fiducial model. The redshifts and global neutral fractions correspond to those of the maps shown in Figure~\ref{Tb_maps}. The shaded region shows the wavenumbers above which the power spectra are unreliable due to aliasing effects in our simulation.}
\label{21cm_PS_Fiducial}
\end{figure}

The evolution in the dimensional 21-cm power spectra for our fiducial model is shown in Figure~\ref{21cm_PS_Fiducial}. The redshifts and global neutral fractions shown correspond to those of the maps in Figure~\ref{Tb_maps}. At early times, when the simulation volume is fully neutral, the 21-cm power spectrum reflects that of the underlying density field (the $z = 13.9$ curve in Figure~\ref{21cm_PS_Fiducial}). Soon after the first regions ionize, the 21-cm and density power spectra diverge and on larger scales ($k \lsim~1$~Mpc$^{-1}$) the 21-cm power spectrum drops briefly (the $z = 9.9$ and 11.1 curves in Figure~\ref{21cm_PS_Fiducial}). This is due to what \cite{LIDZ2008} describe as an `equilibration phase', when over-dense and under-dense regions have similar brightness temperatures due to the `inside-out' nature of reionization. As reionization progresses, ionized regions grow resulting in an increase in large-scale power (small $k$) and a supression of small-scale power. Power on scales corresponding to $k \sim$ 0.1--1~Mpc$^{-1}$ tends to a maximum and the slope of the power spectrum flattens when reionization approaches its midpoint at $\bar{x}_{\rm H\textsc{i}} \approx 0.5$ (which occurs at $z \approx 8$ for our fiducial model). Past this point power on all scales falls until there is no signal. The shaded region in Figure~\ref{21cm_PS_Fiducial} corresponds to wavenumbers $k > 0.7 k_{\rm Ny}$, where $k_{\rm Ny}$ is the Nyquist frequency of our simulation. Due to aliasing effects in our simulation (both in the mass-assignment scheme used to construct the \emph{Tiamat} density grids used by \Meraxes, as well as its regridding within \Meraxes), power for modes in this region are unreliable \citep[see, e.g.,][]{CUI2008}. However, based on fiducial instrumental specifications, current- and future-generation radio interferometers will be most sensitive to modes with $k \sim 0.1$--1~Mpc$^{-1}$ \citep[see, e.g.,][]{LIU2014,PRITCHARD2015}, therefore these aliasing effects do not affect our predictions for observable differentiation between models.

\begin{figure}
\includegraphics[width= 8.5cm]{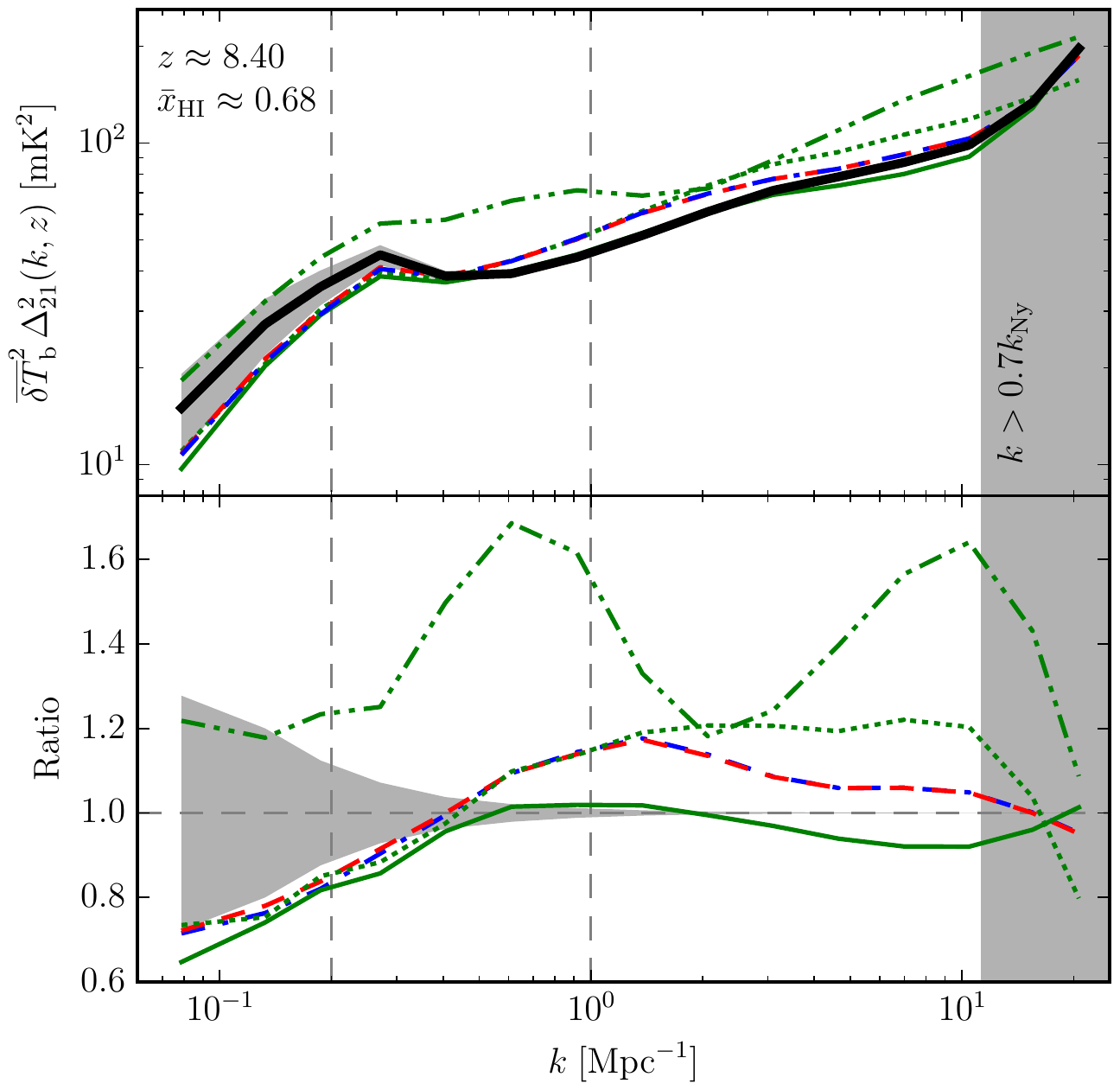} 
\caption{Top panel: Dimensional 21-cm power spectra at $(z, \bar{x}_{\rm H\textsc{i}}) \approx (8.4, 0.68)$. Bottom panel: Ratio of the above power spectra with respect to that of the fiducial model. The vertical dashed lines mark the wavenumber interval where we have compared our models. The shaded regions correspond to the Poisson error for the fiducial model, due to fewer modes available on the largest scales. The shaded region at $k > 0.7 k_{\rm Ny}$ shows the wavenumbers above which the power spectra are unreliable due to aliasing effects in our simulation.}
\label{21cm_PS}
\end{figure}

Figure~\ref{21cm_PS} shows the dimensional 21-cm power spectra for our models at $z \approx 8.4$ ($\bar{x}_{\rm H\textsc{i}} \approx 0.68$) where they have been matched (top panel), together with their ratios with respect to the fiducial model (bottom panel). The shaded regions show the Poisson errors for the fiducial model, due to fewer modes available on the largest scales. The metric we use to compare the power spectra of our models is the maximum fractional difference in dimensional 21-cm power with respect to the fiducial model\footnote{i.e. maximum $\left|\frac{(\overline{\delta T}_{\rm b}^2 \Delta^2_{21})|_{\rm model}}{(\overline{\delta T}_{\rm b}^2 \Delta^2_{21})|_{\rm F}}-1\right|$} in the wavenumber interval $k = 0.2$--1~Mpc$^{-1}$. This interval approximately corresponds to the decade of $k$-modes probed by current- and future-generation radio interferometers, but avoids the small-$k$ values where the Poisson error is relatively large. Using this metric, our results show there is a difference of up to: i) $\approx 17$~per~cent without SNe feedback; ii) $\approx 15$~per~cent without SNe or reionization feedback; iii) $\approx 18$~per~cent using a constant stellar-to-halo mass relationship; iv) $\approx 15$~per~cent using a constant stellar-to-halo mass relationship including only sources with haloes of mass greater than $10^9$~M$_\odot$, and; v) $\approx 69$~per~cent using a constant stellar-to-halo mass relationship including only sources with haloes of mass greater than $10^{10}$~M$_\odot$. These differences are qualitatively similar to those seen in the results for the average size of ionized regions in Section~\ref{Reionization morphology}.

\begin{figure*}
\includegraphics[width= 17.6cm]{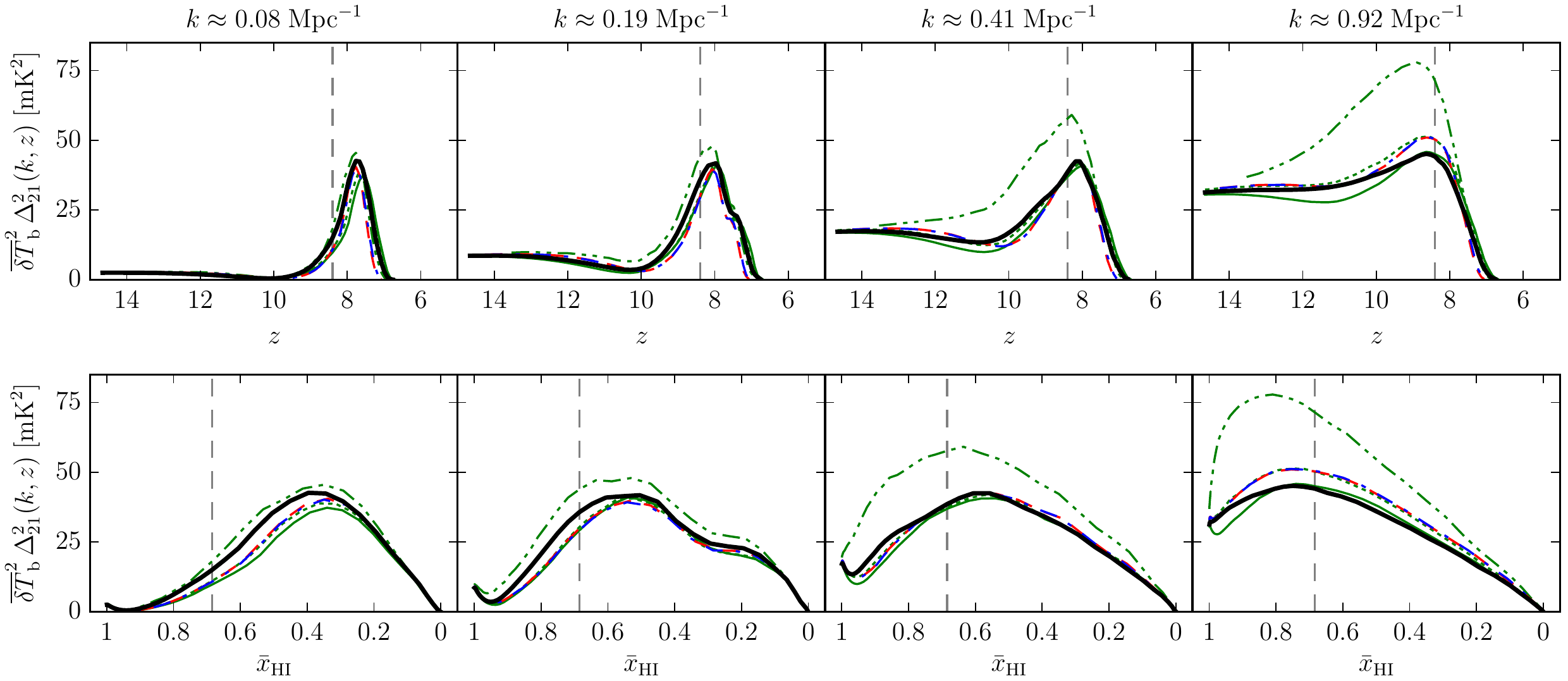}
\caption{Evolution of the dimensional power spectra of 21-cm brightness temperature fluctuations as a function of redshift (upper panels) and global neutral fraction (lower panels). Results are shown for wavenumbers $k \approx 0.08$, 0.19, 0.41 and 0.92~Mpc$^{-1}$. The vertical dashed lines show the redshift and global neutral fractions at which the models have been matched. The same line styles as given in Table~\ref{sim_summary_table} have been used.}
\label{21cm_PS_z}
\end{figure*}

The evolution of spatial fluctuations in the 21-cm signal has been shown to be sensitive to the nature of the sources driving reionization and to the size of their populations \citep[see, e.g.,][]{MLZD2007}. Observing how the 21-cm power spectrum evolves may therefore be used to probe the clustering characteristics and bias of the sources responsible for reionization. In order to explore this we show the evolution of our models' dimensional power spectra for specific wavenumbers ($k \approx 0.08$, 0.19, 0.41 and 0.92~Mpc$^{-1}$) in Figure~\ref{21cm_PS_z}. The top panels show evolution as a function of redshift while the bottom panels show evolution as a function of global neutral fraction. Note that the results at global neutral fractions away from when the models are matched are not as directly comparable on account of relatively small differences in their underlying density fields (due to the model tuning discussed early in Section~\ref{Simulations}). It can be seen that while each model exhibits similar evolutionary behaviour on large scales ($k \lsim\ 0.2$~Mpc$^{-1}$), there are noticable differences on smaller scales.

In particular, not all models' 21-cm power spectra pass through a local minimum during the first half of reionization (i.e. before the maximum). These minima are a feature of models with relatively weakly biased ionizing sources. The CSHR and CSHR.Mcut.10 models provide useful examples to help explain this. The CSHR model includes ionization due to sources associated with haloes of any mass resolved by \emph{Tiamat}, in contrast with the CSHR.Mcut.10 model which has fewer sources hosted by smaller-to-medium-mass haloes. During the early stages of reionization, ionized regions form around these `low-mass sources' in the CSHR model, reducing 21-cm power on scales comparable to the source separation. This leads to persistant minima for $k \lsim\ 1$~Mpc$^{-1}$. However, when only `high-mass sources' are included in the CSHR.Mcut.10 mass-cut model, ionized regions only form about the most clustered, highly biased haloes, leaving 21-cm power on other scales. This leads to the absence of a local minimum in the power spectrum evolution for this model on scales $k \gsim\ 0.4$~Mpc$^{-1}$. This behaviour has been investigated previously \citep[see, e.g.,][]{MLZD2007, LIDZ2008} using models that assume ionizing source luminosities that are proportional to their host halo mass and implementing different source-luminosity relationships. However, these models do not capture the complex interplay between galaxies and the ionization state of the IGM through the effects of feedback (see the discussion in Section~\ref{Discussion}). We find that on smaller scales, semi-analytic modelling of galaxies and feedback reduces the signature of the equilibration phase that is seen in constant stellar-to-halo mass ratio models.

\begin{figure*}
\includegraphics[width= 17.6cm]{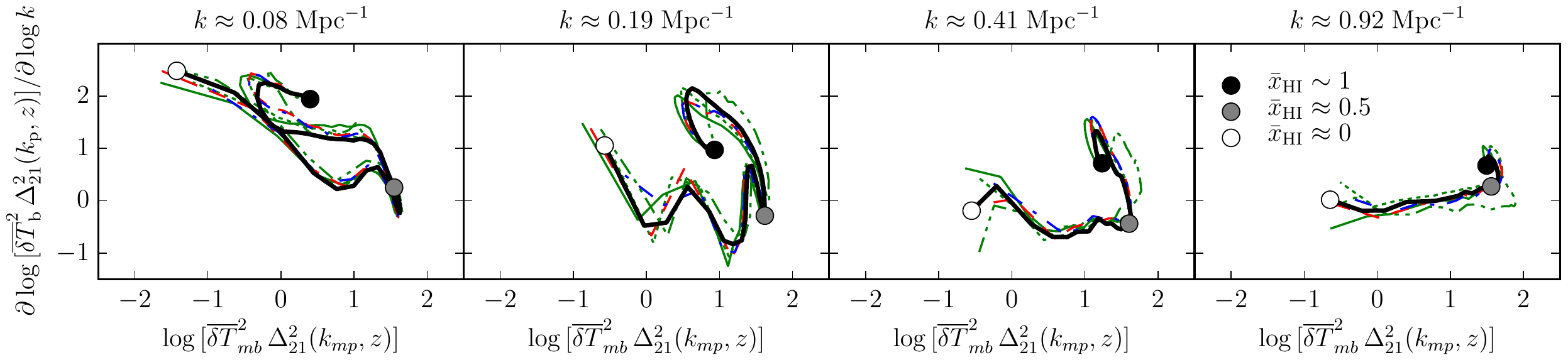}
\caption{Evolution in the gradient of the power spectra of 21-cm brightness temperature fluctuations. Loci are shown for wavenumbers $k \approx 0.08$, 0.19, 0.41 and 0.92~Mpc$^{-1}$. Direction of progression for the fiducial model is indicated by the circles (filled at the start of reionization, empty at the end of reionization). The same line styles  as given in Table~\ref{sim_summary_table} have been used.}
\label{21cm_PS_gradient_evolution}
\end{figure*}

In addition to the 21-cm power spectrum amplitude, the evolutionary behaviour of its slope also serves as a diagnostic of reionization. Since the global neutral fraction is not directly measurable, we follow the work of \cite{LIDZ2008} and \cite{KIM2013a,KIM2013b} by exploring the joint evolution of the 21-cm power spectrum amplitude and its slope at selected scales. Figure~\ref{21cm_PS_gradient_evolution} shows loci of the power spectrum gradient versus amplitude for our models. As was shown in Figure~\ref{21cm_PS_z} for the amplitude, we find that the models exhibit similar evolutionary behaviour during the first half of the reionization process. As before, the main exception is the CSHR.Mcut.10 model. We find a smaller level of difference between models with and without supernova feedback than described in \cite{KIM2013a,KIM2013b}. This is most likely due to the high temporal resolution of our simulation as mentioned in Section~\ref{Reionization morphology}.

\section{Discussion}
\label{Discussion}

As mentioned in Section~\ref{Reionization morphology}, there is a remarkable similarity between the ionization fields of the no-feedback and CSHR models despite the relatively large difference in the host halo mass scale of sources which dominate their reionization. This motivated the implementation of the CSHR mass-cut models, of which, CSHR.Mcut.9 is the most similar to the no-feedback models. As explained, this follows from the fact that their dominant ionizing sources are hosted by haloes in the same mass range. This is in contrast to the CSHR.Mcut.10 model which has a significantly different ionization field owing to the bias of its ionizing sources. While this may not be surprising, since none of these models include feedback effects, what is surprising is the similarity between either the CSHR or CSHR.Mcut.9 model and the fiducial model (which does include feedback effects). This suggests the existance of a mass cut for the CSHR model which best replicates the morphology and 21-cm power spectra results of the fiducial model.

Considering the added complexity of implementing a SAM-based prescription to simulate reionization, such `simple' semi-numerical prescriptions have their utility. While reionization morphology at fixed neutral fraction is not very sensitive to the details of galaxy formation (as shown in Sections~\ref{Reionization morphology} and \ref{21cm power spectra}), it is sensitive to the mass of the haloes which host the bulk of the ionizing sources. Simple EoR parametrizations may therefore be used to predict reionization morphology, and from that infer the dominant ionizing source population. This utility, however, has its limitations. By not including realistic galaxy-formation physics in their formultion, these models cannot be used to investigate the effect of reionization on galaxy formation. Therefore, unlocking the details of galaxy formation requires interpretation with SAMs, in combination with other observations.

\section{Summary and conclusions}
\label{Summary and conclusions}

This work has investigated the dependence of the morphology and statistics of \HIItext~regions on galaxy-formation physics during the Epoch of Reioinization using the Dark-ages, Reionization And Galaxy-formation Observables from Numerical Simulations (DRAGONS). DRAGONS includes the galaxy properties modelled by the semi-analytic model \Meraxes~and includes calculation of the inhomogeneous ionizing UV background using the \tocf~algorithm, providing a self-consistent realization of reionization. This allows us to explore the effect of environment on galaxy formation, subsequent reionization of the IGM, and its observable signatures. We use this coupled model to make updated cosmic 21-cm signal simulations, and predictions for low-frequency 21-cm experiments.

We have presented results for a range of models that capture important physical mechanisms such as reionization and supernova feedback. We have demonstrated that the morphology and statistics of reionization are sensitive to both the ionizing source population and to feedback effects. We have shown how galaxy formation can modify the observable morphology of reionization by looking at the sizes of ionized regions and 21-cm power spectra. Of the galaxy physics we have investigated, we find that supernova feedback plays the most important role in reionzation, and that in the absence of this feedback, \HIItext~regions are up to $\approx 20$~per~cent smaller, while the fractional difference in amplitude of power spectra is up to $\approx 17$~per~cent at fixed ionized fraction. We have compared our SAM-based reionization models with past calculations that assume constant halo mass-to-luminosity ratios. We find that the correct choice of minimum halo mass in these models leads to a reionization morphology that mimics that of a realistic galaxy-formation model. Therefore, reionization morphology at fixed neutral fraction is not uniquely predicted by the details of galaxy-formation physics. However, morphology is sensitive to the mass of the haloes which host the bulk of the reionizing sources. Simple EoR parametrizations therefore have utility for predicting the cosmic 21-cm signal, however, a better understanding of galaxy-formation physics using future 21-cm observations requires interpretation including a model of galaxy formation, in combination with other observations.

In future work we will include a number of improvements to our modelling, including redshift and mass dependency of the escape fraction of ionizing photons, the effect of gas temperature and X-rays, a Markov chain Monte Carlo exploration of the model parameter space, and inhomogenous recombinations.

\section*{Acknowledgments}

This work was supported by the Victorian Life Sciences Computation Initiative (VLSCI), grant reference UOM0005, on its Peak Computing Facility hosted at The University of Melbourne, an initiative of the Victorian Government. Part of this work was performed on the gSTAR national facility at Swinburne University of Technology. gSTAR is funded by Swinburne University of Technology and the Australian Governments Education Investment Fund. AM acknowledges support from the European Research Council (ERC) under the European Unions Horizon 2020 research and innovation program (grant agreement No 638809 AIDA). This research program is funded by the Australian Research Council through the ARC Laureate Fellowship FL110100072 awarded to JSBW. Finally. we would like to thank the anonymous referee for their constructive comments which helped to improve the paper.

\bibliographystyle{pasa-mnras}
\bibliography{mybib}

\newpage

\appendix

\section{Spatial convergence}
\label{App:CT}

In this appendix we demonstrate the spatial convergence of our simulations through refinement of the grids used by \tocf. This is done for both the reionization history and 21-cm power spectra of our fiducial model.

\begin{figure}
\includegraphics[width= 8.5cm]{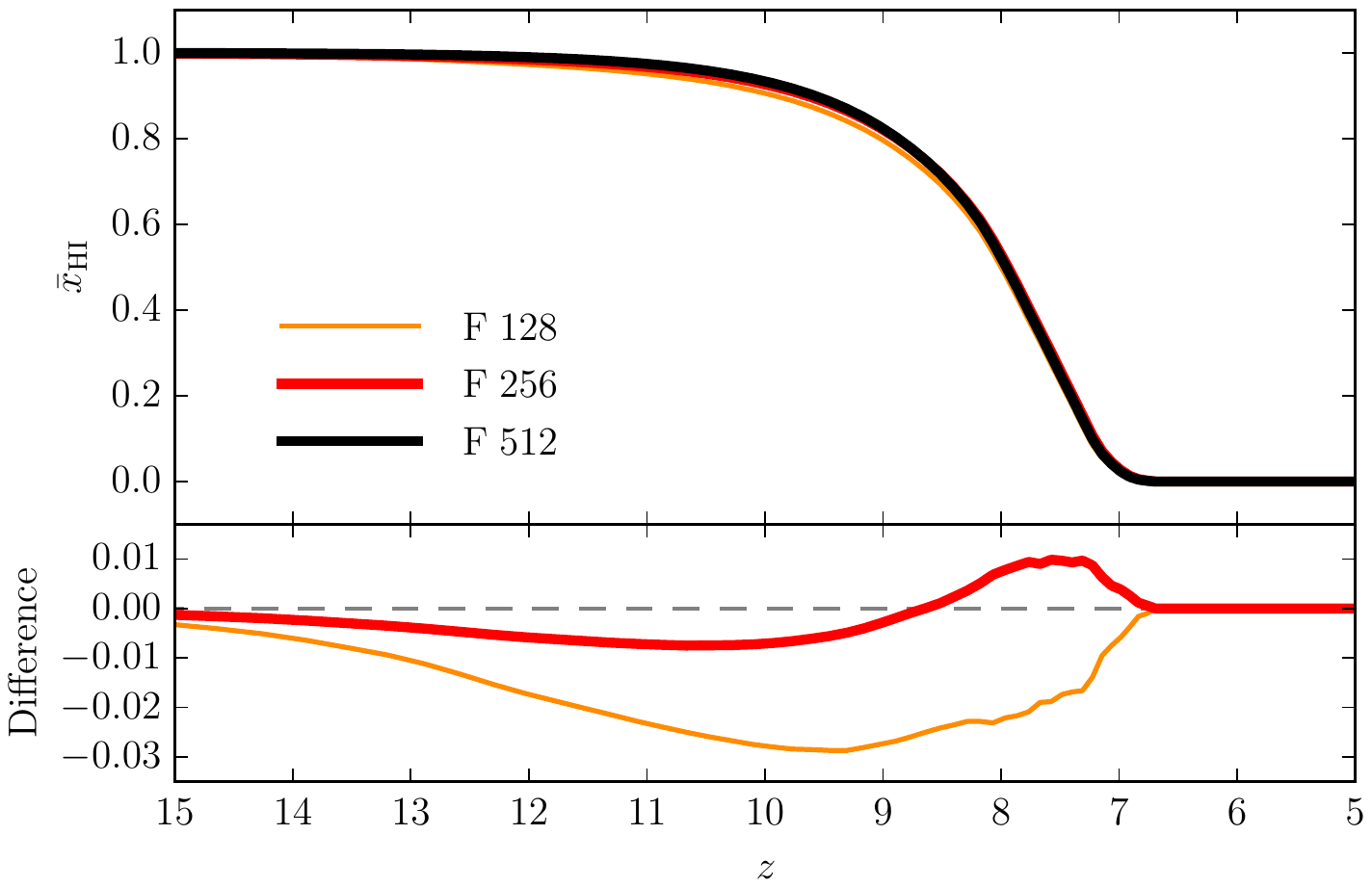}
\caption{Top panel: Globally-averaged neutral fraction, $\bar{x}_{\rm H\textsc{i}}$ as a function of redshift, $z$, for our fiducial model using a $128^3$, $256^3$ and $512^3$ \tocf~grid. Bottom panel: The differences between the globally-averaged neutral fraction of each simulation with respect to the 512$^3$ simulation.}
\label{CT_xH}
\end{figure}

Figure~\ref{CT_xH} shows the globally-averaged neutral fraction, $\bar{x}_{\rm H\textsc{i}}$, as a function of redshift, $z$, for our fiducial model using a $128^3$, $256^3$ and $512^3$ \tocf~grid and the differences between them. We find that our final refinement results in convergence to within $\pm 0.01$ neutral fraction at all redshifts.

\begin{figure}
\includegraphics[width= 8.5cm]{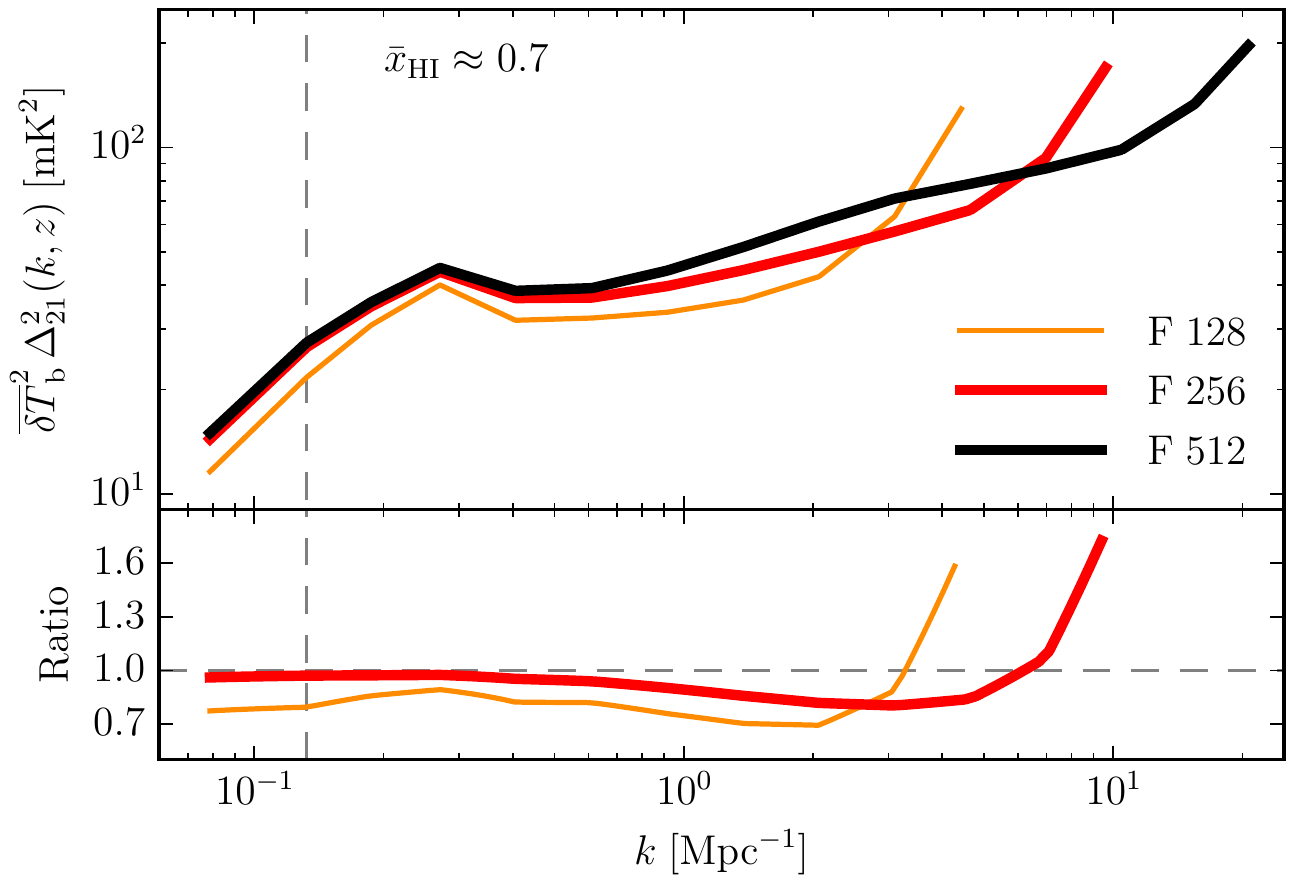}
\caption{Top panel: The dimensional 21-cm power spectra for our fiducial models at $\bar{x}_{\rm H\textsc{i}} \approx 0.68$ using a $128^3$, $256^3$ and $512^3$ \tocf~grid. Bottom panel: The ratio between the dimensional 21-cm power spectra of each simulation with respect to the 512$^3$ simulation. The vertical dotted line shows the scale at which current- and future-generation radio interferometers are most sensitive, which has been used to compare the power spectra of models in this work.}
\label{CT_PS}
\end{figure}

In order to show spatial convergence of the power spectrum, Figure~\ref{CT_PS} shows the dimensional 21-cm power spectra for our fiducial model at $\bar{x}_{\rm H\textsc{i}} \approx 0.68$ using a $128^3$, $256^3$ and $512^3$ \tocf~grid and their ratios with respect to the 512$^3$ simulation. The vertical dotted line shows the scale at which current- and future-generation radio interferometers are most sensitive, which has been used to compare the power spectra of models in this work. We find that successive refinement results in convergence to within $3$~per~cent on this scale.

\end{document}